\pdfoutput=1
\documentclass[10pt,twocolumn,letterpaper]{article}

\usepackage[letterpaper,hmargin=0.67in,top=0.75in,bottom=0.9in,columnsep=0.2in]{geometry}
\usepackage[T1]{fontenc}
\usepackage{graphicx}
\usepackage{subcaption}
\usepackage{cite}
\usepackage{amsmath,amssymb,amsfonts}
\usepackage{mathrsfs}
\usepackage{algorithmicx}
\usepackage{textcomp}
\usepackage{xcolor}
\usepackage{booktabs}
\usepackage{array}
\usepackage[ruled,lined]{algorithm2e}
\usepackage{multirow}
\usepackage{upgreek}
\usepackage{tikz}
\usetikzlibrary{fit,backgrounds}
\usetikzlibrary{shapes,arrows,arrows.meta,calc}
\usepackage{verbatim}
\usepackage{tikz-network}
\usepackage{pgfplots}
\pgfplotsset{compat=1.18}
\usepackage{bm}
\usepackage{bbold}
\usepackage{amsthm}
\usepackage{hyperref}
\hypersetup{
    colorlinks=true,
    linkcolor=blue,
    anchorcolor=blue,
    citecolor=blue,
    pdftitle={Transversal Passivity: Dissipation Relative to Input and Output Manifolds},
    pdfauthor={Feng-Yu Yue, Daniel Zelazo}
}

\newtheorem{assumption}{Assumption}
\newtheorem{proposition}{Proposition}
\newtheorem{lemma}{Lemma}
\newtheorem{theorem}{Theorem}
\newtheorem{definition}{Definition}
\newtheorem{corollary}{Corollary}

\newtheorem{example}{Example}
\newtheorem{remark}{Remark}

\newcommand{\R}{\mathbb{R}}

\newcommand{\dist}{\operatorname{dist}}

\providecommand{\IEEEPARstart}[2]{#1#2}

\title{\bfseries Transversal Passivity: Dissipation Relative to\\ Input and Output Manifolds}
\author{%
  Feng-Yu Yue \qquad Daniel Zelazo\\[0.5ex]
  {\normalsize The Stephen B. Klein Faculty of Aerospace Engineering,}\\
  {\normalsize Technion--Israel Institute of Technology, Haifa, Israel}\\
  {\normalsize\ttfamily fengyu.yue@campus.technion.ac.il, dzelazo@technion.ac.il}%
}
\date{}

\begin{document}

\makeatletter
\twocolumn[
\begin{@twocolumnfalse}
\maketitle
\vspace{-1.5em}
\begin{abstract}
This paper introduces transversal passivity, a local dissipation property for regulating input and output signals to prescribed manifolds. Its supply rate is built from the transverse port errors induced by nearest-point projections, so that motion along the manifolds is not penalized. We give a necessary condition and a constructive sufficient condition for the property and relate it to existing passivity notions. For feedback interconnections through negative-adjoint linear channels, power conservation no longer cancels the transverse supply. The residual is governed by compatibility defects that depend only on the channels and the prescribed manifolds, so it can be assessed before any closed-loop analysis, and it falls into a hierarchy of exact compatibility, quadratic cancellation, and nonzero quadratic residuals. Since transversal storages are only positive semidefinite, port convergence is obtained by combining compact confinement, for instance by conserved quantities, with a dissipation margin, and we identify a geometric obstruction under which no quadratic margin is available near the feasible states. A federated learning case study with heterogeneous data shows that port convergence can be certified even though no server target is compatible with the interconnection, and that the obstruction arises for a natural choice of target.

\end{abstract}
\vspace{1.5em}
\end{@twocolumnfalse}
]
\makeatother

\section{Introduction} 

\IEEEPARstart{W}{e} consider control problems in which input and output signals are required to approach prescribed manifolds, rather than prescribed points. Such problems arise in many applications, including path-following \cite{yao2021pathfollowing}, control allocation \cite{johansen2013allocation}, formation control \cite{dorfler2010formation}, synchronization \cite{sarlette2009consensusoptimization}, and the training of overparameterized learning models, whose interpolating solutions form a manifold \cite{liu2022loss}. 
This class of problems leaves a geometric degree of freedom in the control design -- we need only penalize the transverse deviations of the signals from the target manifolds while leaving tangential motion along the manifolds unconstrained. This motivates an input-output formulation, where the deviations are measured without imposing a tangential reference. 

Passivity theory provides a natural starting point for such a formulation. A passive system is characterized by a storage function whose derivative is bounded from above by a supply rate defined on the system inputs and outputs. This input-output characterization relates passivity to Lyapunov theory and finite-gain input--output stability~\cite{Khalil2008Nonlinear}. A further advantage is compositionality, which guarantees that the feedback interconnection of two passive systems is passive, and the sum of the individual storage functions is a storage function of the interconnection~\cite{willems1972dissipative}.

Existing passivity variants reference the supply rate to points or trajectories; shifted passivity, for instance, uses a single input-output pair, which is not available a priori when the targets form a manifold and the limit depends on the initial condition. To adapt passivity to manifold-target problems, we introduce a new notion termed \emph{transversal passivity}, which is a passivity property defined locally with respect to prescribed input and output manifolds. Its storage function is continuously differentiable and positive semidefinite, and its supply rate measures only the finite transverse errors induced by nearest-point projections of the inputs and outputs onto these manifolds.

Some related approaches retain this separation in the state space, but represent and constrain the transverse deviations differently. Transverse contraction~\cite{manchester2014transversecontraction} imposed contraction on infinitesimal state variations transverse to the system flow. Transverse exponential stability~\cite{andrieu2016transversestability} required exponential decay of the transverse state coordinates about an invariant manifold. Local transverse feedback linearization~\cite{nielsen2008transverselinearization} used local coordinates and regular static feedback so that the dynamics transverse to a controlled-invariant manifold became linear and controllable. Immersion and invariance~\cite{astolfi2003immersion} and nonlinear output regulation~\cite{isidori1990output} also target invariant manifolds, but as design methods for a given closed loop: they construct a controller that renders a manifold invariant and attractive, rather than a port property of the subsystems that is preserved under interconnection. Closer in spirit, passivity of input--output relations with respect to submanifolds~\cite{Max2017submanifold} measures the distances of the input and output signals to prescribed submanifolds, but only at the signal level, without a storage function, and for the standard negative feedback interconnection.

We study the symmetric feedback interconnection of two transversally passive systems through linear channels, in which the reverse channel is the negative adjoint of the forward channel; standard negative feedback is the special case in which the forward channel is the identity. The interconnection is power-conserving~\cite{arcak2007passivity}, i.e., the net power through the ports vanishes, and this cancellation underpins passivity-based closed-loop analysis~\cite{van2013pHgraph,burger2014duality}, including output synchronization~\cite{chopra2006passivity} and consensus on directed graphs~\cite{yue2025Spassivity,yue2025EJC}. In these analyses the supply rates are referenced to points or to the other agents, so the target set enters only through the storage or the interconnection. The first difficulty is that, for transversal passivity, the relevant inner product is instead that of the transverse port errors, which power conservation does not force to vanish. A second difficulty is that transversal storages are local and only positive semidefinite, so their sublevel sets need not be compact and dissipation alone does not yield convergence of the ports; related issues arise in the passivity-based stabilization of noncompact sets, studied through reduction principles and set detectability~\cite{el2010reduction}.

The paper makes three contributions. (i) We introduce transversal passivity, give a necessary condition and a constructive sufficient condition for it, and show that it recovers shifted and incremental passivity as special cases and connects to differential passivity at equilibria. (ii) We show that, under negative-adjoint feedback, transversal passivity composes up to a residual supply. The residual is governed by compatibility defects, which measure whether the channels preserve the transverse geometry and depend only on the channels and the prescribed manifolds, so it can be checked before any closed-loop analysis. We classify this residual into three cases: exact compatibility; quadratic cancellation, in which the channels need not preserve the transverse geometry but the residual vanishes to second order; and a nonzero quadratic residual. (iii) We give conditions for port convergence that combine compact confinement, for instance from a conserved quantity whose level set is fixed by the initial condition, with a dissipation margin, and we identify a geometric obstruction under which no quadratic margin is available near the feasible states. A federated learning case study with heterogeneous client data shows that these results certify port convergence even though no server target is compatible with the interconnection, and that the obstruction arises for a natural choice of target.


The remainder of this paper is organized as follows. Section~\ref{sec:preliminaries} collects preliminaries, Section~\ref{sec:TP_def} develops transversal passivity, Section~\ref{sec:linear_feedback} treats feedback interconnection and the residual supply, Section~\ref{sec:linear_analysis} establishes port convergence, Section~\ref{sec:fl} presents the federated learning case study, and Section~\ref{sec:conclusion} concludes the paper.

\subsubsection*{Notation}
$\mathbb{R}_{\ge 0}$ and $\mathbb{R}_{>0}$ denote the nonnegative and positive reals, $\langle\cdot,\cdot\rangle$ the Euclidean inner product, and $\|\cdot\|$ the induced norm; for linear and multilinear maps, $\|\cdot\|$ is the induced operator norm. A product space carries the norm $\|(a_1,a_2)\|=(\|a_1\|^2+\|a_2\|^2)^{1/2}$. We write $I_n$ for the $n\times n$ identity, $\mathbb 1_N$ for the vector of $N$ ones, $\otimes$ for the Kronecker product, $A^\top$ for the transpose, and $\ker A$ for the kernel of a matrix $A$; $\operatorname{span}$ denotes linear span. For a symmetric matrix $P$, $P\succ0$ ($P\succeq0$) means $P$ is positive definite (semidefinite). The unit sphere in $\mathbb R^{n}$ is $\mathbb S^{n-1}$, and $\operatorname{id}$ is the identity map. For a linear subspace $M\subseteq\mathbb{R}^n$, $\mathrm{P}_M$ is the orthogonal projection onto $M$. For a set $S$, $\overline S$ is its closure, $\dist(x,S):=\inf_{y\in S}\|x-y\|$ is the distance from $x$ to $S$, and $h^{-1}(S)$ is the preimage of $S$ under a map $h$, which need not be invertible. $C^k$ denotes the class of $k$ times continuously differentiable maps. If $\rho$ is differentiable at $x$, its differential is the linear map $\mathrm{d}_x\rho$, acting on $v$ as $\mathrm{d}_x\rho[v]$; if $\rho$ is twice differentiable, $\mathrm d_x^2\rho$ is the symmetric bilinear second differential, acting as $\mathrm d_x^2\rho[v,w]$. For a scalar function $V$, $\nabla V$ and $\nabla^2V$ denote its gradient and Hessian. Landau symbols $o(\cdot)$ and $O(\cdot)$ refer to the limit of their argument tending to zero.

\section{Preliminaries}\label{sec:preliminaries}
This section introduces the definitions and background material used throughout the paper, including notions from passivity theory and differential geometry.
\subsection{Standard Passivity}\label{ssec:passivity}
Consider $\dot x=f(x,u)$, $y=h(x,u)$, with $x\in\mathbb R^n$ and $u,y\in\mathbb R^m$. A passivity inequality is certified by a continuously differentiable storage $Q:\mathbb R^n\to\mathbb R_{\geq0}$ and indices $\varepsilon,\gamma\geq0$ satisfying
\begin{equation}\label{eq:standard_passivity}
\mathrm d_xQ[f(x,u)]\leq u^\top y-\varepsilon\|y\|^2-\gamma\|u\|^2
\end{equation}
for every $(x,u)$ \cite{Khalil2008Nonlinear}. The system is \emph{passive} if $\varepsilon=\gamma=0$, \emph{output-strictly passive} if $\varepsilon>0$ and $\gamma=0$, and \emph{input-strictly passive} if $\varepsilon=0$ and $\gamma>0$. The right-hand side of \eqref{eq:standard_passivity} is called the supply rate \cite{willems1972dissipative}. 
We will discuss other passivity variants in Subsection~\ref{ssec:vs_standard_passivity}.

\subsection{Tubular Neighborhoods and Nearest-Point Projections}\label{ssec:tubular_retraction}
We employ basic notions from differential geometry in this work.  Let $\mathcal{M}$ be a smoothly embedded submanifold of $\mathbb{R}^n$. Denote its tangent space at $p\in\mathcal M$ by $T_p\mathcal M$ and its normal space by $N_p\mathcal M:=(T_p\mathcal M)^\perp$. 

For a continuous positive function $\epsilon:\mathcal M\to\mathbb R_{>0}$, define the \emph{$\epsilon$-neighborhood} of $\mathcal M$ by \cite{guillemin2010differential}
$$ \mathcal N:=\left\{ z\in\mathbb R^n: \|z-p\|<\epsilon(p) \text{ for some }p\in\mathcal M \right\}. $$
The $\epsilon$-neighborhood theorem \cite[Chapter 2]{guillemin2010differential} guarantees that $\epsilon$ can be chosen so that every $z\in\mathcal N$ has a unique nearest point on $\mathcal M$. If $\mathcal M$ is compact, then $\epsilon$ can be taken as a positive constant. Therefore, the associated nearest-point projection
\begin{equation}
  r:\mathcal N\to\mathcal M, \quad r(z):=\operatorname*{arg\,min}_{p\in\mathcal M}\|z-p\|^2,
\end{equation}
is well-defined and smooth. For every $z\in\mathcal N$, define the transverse error map by
\begin{equation}\label{eq:manifold_error} 
e(z):=z-r(z).
\end{equation}
Then $z=r(z)+e(z)$ and $e(z)\in N_{r(z)}\mathcal M$ is the normal displacement of $z$ from its base point $r(z)\in\mathcal M$. 
Moreover, the nearest-point property gives
\begin{align}\label{eq:emap_distance}
\|e(z)\|=\|z-r(z)\|=\dist(z,\mathcal M).
\end{align}

With $\epsilon$ chosen as above, $\mathcal N$ is a tubular neighborhood of $\mathcal M$ and $r$ is the associated smooth orthogonal tubular retraction \cite{lee2003smoothmanifolds}; we use the two terms interchangeably.

Now, we summarize some standard properties of the nearest-point projection, which are fundamental to the subsequent discussion.
\begin{lemma}[\cite{leobacher2021regularity}]\label{lem:tubular_projection_identities}
Consider a submanifold $\mathcal{M}$, tubular neighborhood $\mathcal{N}$, nearest-point projection $r$, and transverse error map $e$. Define $V_{\mathcal{M}}(z) = \tfrac12\|e(z)\|^2$. Then the following hold:
\begin{enumerate}
\item[(i)] For every $z\in\mathcal{M}$, the projection fixes $z$ and the transverse error vanishes,
\begin{equation}
    r(z)=z, \quad e(z)=0.
\end{equation}
Moreover, the corresponding differentials of $r$ and $e$ satisfy
  \begin{equation}
    \mathrm{d}_z r = \mathrm{P}_{T_z\mathcal{M}}, \text{ and }
    \mathrm{d}_z e = \mathrm{P}_{N_z\mathcal{M}}.
  \end{equation}
\item[(ii)] For every $z\in\mathcal{N}$ and $v\in \R^n$, the derivative of $V_{\mathcal{M}}$ at $z$ in the direction $v$ is given by
  \begin{equation}
    \mathrm{d}_z V_{\mathcal{M}}[v] = \bigl\langle e(z),\,v\bigr\rangle,
  \end{equation}
  since $\mathrm{d}_z r[v]\in T_{r(z)}\mathcal M$ is orthogonal to $e(z)\in N_{r(z)}\mathcal M$.
\end{enumerate}
\end{lemma}

\section{Transversal Passivity}\label{sec:TP_def}
In this section, we introduce transversal passivity, a state-space certificate for the input and output behavior relative to prescribed manifolds. We then establish a necessary condition and a constructive sufficient condition for this property, and relate it to other established passivity variants.

Consider the nonlinear system,
\begin{equation}\label{eq:dynsys}
    \Sigma:\quad \dot x = f(x,u), \quad y = h(x,u),
\end{equation}
where $x\in \mathcal X\subseteq \mathbb R^n$, $u\in \mathcal U\subseteq \mathbb R^m$, and $y\in \mathcal Y\subseteq \mathbb R^m$ denote the state, input, and output, respectively, of $\Sigma$; we refer to $u$ and $y$ as the ports of $\Sigma$. Here, $\mathcal X$, $\mathcal U$ and $\mathcal Y$ are open sets, and $f:\mathcal X\times \mathcal U\to \mathbb R^n$ and $h:\mathcal X\times \mathcal U\to \mathbb R^m$ are continuously differentiable. 

We are interested in control problems where the input and output signals are required to approach prescribed input manifolds $\mathcal M_u\subseteq\mathcal U$ and output manifolds $\mathcal M_y\subseteq\mathcal Y$. In such problems, the signals are allowed to move persistently along the manifolds, while their deviations from the manifolds are required to decay. This motivates a notion of passivity whose supply rate measures only the transverse errors of these signals.

In this direction, let $\mathcal{M}_u\subseteq \mathcal U$ and $\mathcal{M}_y\subseteq \mathcal Y$ be embedded submanifolds. Following Subsection~\ref{ssec:tubular_retraction}, choose tubular neighborhoods $\mathcal N_u\subseteq\mathcal U$ and $\mathcal N_y\subseteq\mathcal Y$ on which the associated nearest-point projections $r_u, r_y$ are uniquely defined and smooth. The corresponding transverse port error maps are
\begin{equation*}
e_u(u)=u-r_u(u) \text{ and } e_y(y)=y-r_y(y).
\end{equation*}
By~\eqref{eq:emap_distance}, these maps satisfy $\|e_u(u)\|=\dist(u,\mathcal M_u)$ and $\|e_y(y)\|=\dist(y,\mathcal M_y)$.

Since the transverse error maps are defined only on their respective tubular neighborhoods, a state-input pair $(x,u)$ can be considered only when both $u\in\mathcal N_u$ and the corresponding output $h(x,u)\in\mathcal N_y$. This motivates the tubular domain
\begin{equation}\label{eq:tubular-domain}
\mathcal D:=\left\{(x,u)\in\mathcal X\times\mathcal U:u\in\mathcal N_u,\; h(x,u)\in\mathcal N_y
\right\}.
\end{equation}
The domain $\mathcal D$ is open because $h$ is continuous and $\mathcal N_u,\mathcal N_y$ are open.

\begin{figure}[!h]
    \centering
    \includegraphics[width=\linewidth]{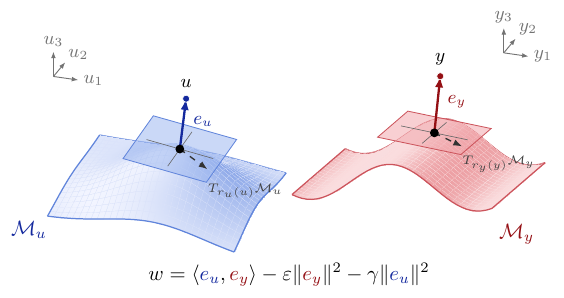}
    \caption{Transversal supply rate. Only the normal displacements $e_u$ and $e_y$ (solid) enter the supply rate; motion along the tangent directions (dashed) is not penalized.}
    \label{fig:supply_rate}
\end{figure}

We now define transversal passivity on the tubular domain $\mathcal D$. This notion provides a state-space certificate for input-output behavior relative to the prescribed manifolds.
\begin{definition}[Transversal Passivity]
\label{def:transversal_passivity}
Consider the system~\eqref{eq:dynsys}, and assume that the tubular domain $\mathcal D$ in~\eqref{eq:tubular-domain} is nonempty. If there exist a continuously differentiable storage function $V:\mathcal X\to\mathbb R_{\geq0}$ and constants $\varepsilon,\gamma\geq0$ such that, for every $(x,u)\in\mathcal D$,
\begin{equation}\label{eq:io_manifold_passivity}
\begin{aligned}
\mathrm{d}_xV[f(x,u)]\leq {}& \langle e_u(u),e_y(h(x,u))\rangle -\varepsilon\|e_y(h(x,u))\|^2 \\
&-\gamma\|e_u(u)\|^2 ,
\end{aligned}
\end{equation}
then the system is called
\begin{itemize}
    \item $(\mathcal M_u,\mathcal M_y)$-passive if
    $\varepsilon=\gamma=0$;
    \item output-strictly $(\mathcal M_u,\mathcal M_y)$-passive
    if $\varepsilon>0$ and $\gamma=0$;
    \item input-strictly $(\mathcal M_u,\mathcal M_y)$-passive
    if $\varepsilon=0$ and $\gamma>0$;
    \item input-output strictly
    $(\mathcal M_u,\mathcal M_y)$-passive if
    $\varepsilon>0$ and $\gamma>0$.
\end{itemize}
\end{definition}
The supply rate constrains only the normal displacements of $u$ and $y$ from $\mathcal M_u$ and $\mathcal M_y$, while motion tangent to the manifolds is not penalized; see Fig.~\ref{fig:supply_rate}. The term \emph{transversal} refers to these normal displacements and this usage is distinct from the classical transversality in differential geometry \cite{lee2003smoothmanifolds}, which concerns non-tangential and non-degenerate intersections of submanifolds. 

\begin{remark}[Locality]
Inequality~\eqref{eq:io_manifold_passivity} and its integral form hold on any time interval over which $(x(t),u(t))$ remains in $\mathcal D$. The definition itself does not ensure that trajectories remain in this domain.
\end{remark}

\begin{remark}
Decreasing $\varepsilon$ or $\gamma$ preserves~\eqref{eq:io_manifold_passivity}, so in the input-output strict case one may always assume $\varepsilon\gamma<1/4$. The coefficient matrix $\left[\begin{smallmatrix}-\gamma&1/2\\1/2&-\varepsilon\end{smallmatrix}\right]$ of the supply rate then has determinant $\varepsilon\gamma-1/4<0$, so the supply rate is indefinite.
\end{remark}


Definition \ref{def:transversal_passivity} gives a dissipation inequality that must hold on the entire tubular domain, leading to two immediate questions:  what does this dissipation inequality imply about the entire system, and how can one find a storage function that satisfies it? The next two subsections address these questions by establishing a necessary condition and a constructive sufficient condition for transversal passivity.

\subsection{A Necessary Condition}
We now consider the infinitesimal consequence of transversal passivity at an equilibrium whose ports lie on the prescribed manifolds.

Let $(x^\star,u^\star)\in\mathcal D$ be an equilibrium pair of \eqref{eq:dynsys}, so that
\begin{equation}\label{eq:sys_equilibrium}
f(x^\star,u^\star)=0,\quad y^\star=h(x^\star,u^\star),
\end{equation}
where $u^\star\in\mathcal M_u$ and $y^\star\in\mathcal M_y$. Assume that $f$ and $h$ are differentiable at $\hspace{-0.05cm}(x^\star,u^\star)\hspace{-0.05cm}$. The linearization of \eqref{eq:dynsys} at $\hspace{-0.05cm}(x^\star,u^\star)\hspace{-0.05cm}$ is
\begin{equation}\label{eq:sys_linearization}
\delta\dot x=\bar A\delta x+\bar B\delta u,\quad
\delta y=\bar C\delta x+\bar D\delta u,
\end{equation}
where $\bar A=\mathrm{d}_x f[x^\star,u^\star]$, $\bar B=\mathrm{d}_u f[x^\star,u^\star]$, $\bar C=\mathrm{d}_x h[x^\star,u^\star]$, and $\bar D=\mathrm{d}_u h[x^\star,u^\star]$. For brevity, write $\mathrm P_u:=\mathrm P_{N_{u^\star}\mathcal M_u}$ and $\mathrm P_y:=\mathrm P_{N_{y^\star}\mathcal M_y}$ for the orthogonal projections onto the normal spaces of $\mathcal M_u$ at $u^\star$ and of $\mathcal M_y$ at $y^\star$.

The following proposition gives a necessary local condition for transversal passivity. At the considered equilibrium, transversal passivity of \eqref{eq:dynsys} implies that the linearization \eqref{eq:sys_linearization} satisfies the standard passivity inequality in the transverse components of infinitesimal $\delta u$ and $\delta y$.

\begin{proposition}\label{prop:necessary_condition}
Consider the equilibrium $(x^\star,u^\star)\in\mathcal D$ with $u^\star\in\mathcal M_u$ and $y^\star\in\mathcal M_y$, and the associated linearized system \eqref{eq:sys_linearization}. Suppose that \eqref{eq:io_manifold_passivity} holds on $\mathcal D$ with a storage function $V$ that is $C^2$ near $x^\star$ and has a local minimum at $x^\star$. Set $Q=\nabla^2V(x^\star)$ and $\delta V(\delta x)=\frac12\delta x^\top Q\delta x$. Then $Q\succeq0$, and along \eqref{eq:sys_linearization},
\begin{equation}\label{eq:necessary_TPassivity}
\delta\dot V\leq \langle \mathrm P_u\delta u,\, \mathrm P_y\delta y\rangle
-\varepsilon\|\mathrm P_y\delta y\|^{2}-\gamma\|\mathrm P_u\delta u\|^{2}.
\end{equation}
\end{proposition}
\begin{proof}
Fix $(\delta x,\delta u)$ and set $x_\lambda=x^\star+\lambda\delta x$, $u_\lambda=u^\star+\lambda\delta u$; since $\mathcal D$ is open and contains $(x^\star,u^\star)$, these pairs lie in $\mathcal D$ for all sufficiently small $\lambda>0$. The local minimum gives $\nabla V(x^\star)=0$ and $Q\succeq0$, so $\nabla V(x_\lambda)=\lambda Q\delta x+o(\lambda)$, while $f(x^\star,u^\star)=0$ gives $f(x_\lambda,u_\lambda)=\lambda(\bar A\delta x+\bar B\delta u)+o(\lambda)$. Since $u^\star\in\mathcal M_u$ and $y^\star\in\mathcal M_y$, Lemma~\ref{lem:tubular_projection_identities}(i) gives $e_u(u^\star)=0$, $e_y(y^\star)=0$, $\mathrm d_{u^\star}e_u=\mathrm P_u$ and $\mathrm d_{y^\star}e_y=\mathrm P_y$, so that
\begin{equation*}
e_u(u_\lambda)=\lambda \mathrm P_u\delta u+o(\lambda), \quad
e_y\bigl(h(x_\lambda,u_\lambda)\bigr)=\lambda \mathrm P_y\delta y+o(\lambda).
\end{equation*}
Both sides of \eqref{eq:io_manifold_passivity} evaluated at $(x_\lambda,u_\lambda)$ are therefore $O(\lambda^{2})$. Dividing by $\lambda^{2}$ and letting $\lambda\to0$ leaves \eqref{eq:necessary_TPassivity}, since $\delta\dot V=\delta x^\top Q\,\delta\dot x=\delta x^\top Q(\bar A\delta x+\bar B\delta u)$ along \eqref{eq:sys_linearization}.
\end{proof}

\begin{remark}[Matrix form]\label{rem:necessary_lmi}
Since $\mathrm P_u$ and $\mathrm P_y$ are orthogonal projections, the right-hand side of~\eqref{eq:necessary_TPassivity} is the quadratic supply $\tfrac12[\delta y^\top\ \delta u^\top]\Pi\,[\delta y^\top\ \delta u^\top]^\top$ with
$$
\Pi:=\begin{bmatrix}-2\varepsilon\mathrm P_y & \mathrm P_y\mathrm P_u\\ \mathrm P_u\mathrm P_y & -2\gamma\mathrm P_u\end{bmatrix}.
$$
Because~\eqref{eq:necessary_TPassivity} holds for every $(\delta x,\delta u)\in\mathbb R^n\times\mathbb R^m$, it is equivalent to the linear matrix inequality
\begin{equation}\label{eq:necessary_lmi}
\begin{bmatrix}\bar A^\top Q+Q\bar A & Q\bar B\\ \bar B^\top Q & 0\end{bmatrix}
-\begin{bmatrix}\bar C & \bar D\\ 0 & I_m\end{bmatrix}^{\!\top}\Pi
\begin{bmatrix}\bar C & \bar D\\ 0 & I_m\end{bmatrix}\preceq0 .
\end{equation}
For $\mathrm P_u=\mathrm P_y=I_m$, \eqref{eq:necessary_lmi} is the Kalman--Yakubovich--Popov inequality for passivity of the linearization~\cite{willems1972dissipative2}; transversal passivity inserts the normal projections into the supply matrix. Proposition~\ref{prop:necessary_condition} then produces a convex test: if no $Q\succeq0$ satisfies~\eqref{eq:necessary_lmi}, no storage function of the type considered certifies transversal passivity with indices $(\varepsilon,\gamma)$ at this equilibrium. The test is necessary only, since~\eqref{eq:necessary_lmi} constrains the linearization alone.
\end{remark}

This proposition demonstrates that transversal passivity enforces the passivity constraint only along the normal directions of the prescribed manifolds. It also reveals a structural obstruction to input-strict transversal passivity. 
\begin{corollary}[Obstruction to input strictness]\label{cor:input_strictness_obstruction}
Under the assumptions of Proposition~\ref{prop:necessary_condition}, suppose that the linearization has no direct input--output feedthrough, i.e., $\bar D=0$, and that $\mathcal M_u$ has positive codimension at $u^\star$, i.e., $N_{u^\star}\mathcal M_u\neq\{0\}$. Then $\gamma=0$.
\end{corollary}
\begin{proof}
The inequality \eqref{eq:necessary_TPassivity} holds for every pair of directions $(\delta x,\delta u)\in\mathbb R^n\times\mathbb R^m$. Since $N_{u^\star}\mathcal M_u\neq\{0\}$, we may choose $\delta u\in N_{u^\star}\mathcal M_u$ with $\|\delta u\|=1$, and we set $\delta x=0$. Then $\delta \dot{V}=0$, while $\bar D=0$ gives $\delta y=\bar C\delta x+\bar D\delta u=0$ and hence $\mathrm P_y\delta y=0$. Moreover, $\mathrm P_u\delta u=\delta u$, so \eqref{eq:necessary_TPassivity} reduces to $0\leq-\gamma$. Since $\gamma\geq0$ by Definition~\ref{def:transversal_passivity}, we conclude $\gamma=0$.
\end{proof}
Equivalently, input strictness is available only when $N_{u^\star}\mathcal M_u=\{0\}$. In that case $e_u$ vanishes identically near $u^\star$, so the term $\gamma\|e_u\|^2$ is identically zero and imposes no constraint. This is the transversal counterpart of the classical fact that a passive system with relative degree at least one cannot be input-strictly passive~\cite{brogliato2020dissipative}.

\subsection{A Constructive Sufficient Condition}
We next construct a storage candidate from the squared distance of the output to its prescribed manifold, obtaining a sufficient condition that can be checked directly from the system data.

\begin{proposition}[Output-distance storage certificate]
\label{prop:sufficient_tpassivity}
  Consider the system $\Sigma$ in \eqref{eq:dynsys} with $y=h(x)$ and $h\in C^1$. Restrict its state domain to $\mathcal X_y:=h^{-1}(\mathcal N_y)$ and define
  \begin{equation}\label{eq:storage_sufficient}
  V(x):=\frac12\|e_{y}(h(x))\|^2.
  \end{equation}
If, for some $\varepsilon\geq 0$ and every $(x,u)\in\mathcal D$,
\begin{equation}\label{eq:sufficient}
\begin{aligned}
\left\langle e_y(h(x)),\mathrm{d}_xh[f(x,u)]\right\rangle \leq {}&
\left\langle e_u(u),e_y(h(x))\right\rangle \\
&-\varepsilon\|e_y(h(x))\|^2,
\end{aligned}
\end{equation}
then $V$ certifies transversal passivity on $\mathcal D$ with indices $(\varepsilon,\gamma)=(\varepsilon,0)$. In particular, the system is output-strictly $(\mathcal M_u,\mathcal M_y)$-passive if $\varepsilon>0$.
\end{proposition}

\begin{proof}
The function $V$ is continuously differentiable and nonnegative on $\mathcal X_y$. By Lemma~\ref{lem:tubular_projection_identities}(ii) and the chain rule,
$$ \mathrm{d}_xV[f(x,u)] = \left\langle e_y(h(x)),\mathrm{d}_xh[f(x,u)]\right\rangle. $$
Thus, \eqref{eq:sufficient} gives \eqref{eq:io_manifold_passivity} with $\gamma=0$.
\end{proof}
The choice $\gamma=0$ is consistent with Corollary~\ref{cor:input_strictness_obstruction}. Under the hypotheses of the corollary, the absence of direct feedthrough precludes input strictness when $\mathcal M_u$ has positive codimension.

For an input-dependent output $y=h(x,u)$, the squared output distance generally depends on $u$, and its time derivative involves $\dot u$ when the input is differentiable. The construction therefore requires additional assumptions or an augmented-state formulation in that setting.

\begin{figure}[t]
    \centering
    \includegraphics[width=0.5\linewidth]{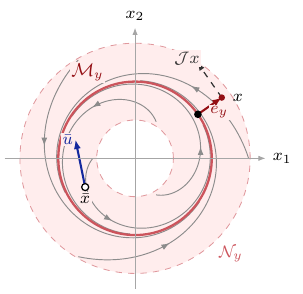}
    \caption{Example~\ref{exm:nonpassive_oscillator}. Unforced trajectories (gray) converge to $\mathcal M_y$ inside the annulus $\mathcal N_y$; the radial error $e_y$ is damped and the rotation $\mathcal Jx$ is unpenalized. At $\bar x$, the input $\bar u$ holds the state at rest with $\bar u^\top\bar y<0$.}
    \label{fig:oscillator}
\end{figure}

The following example shows that transversal passivity can hold even when standard passivity fails.
\begin{example}[A transversally passive oscillator that is not passive]
\label{exm:nonpassive_oscillator}
Consider the nonlinear oscillator
\begin{equation*}
    \dot x=\mathcal Jx-\left(1-\frac{1}{\|x\|}\right)x+u,
    \quad y=x,
\end{equation*}
where $x\in\mathcal X:=\mathbb R^2\setminus\{0\}$, $u\in\mathbb R^2$, and $\mathcal J= \begin{bmatrix}  0 & -1\\  1 & 0 \end{bmatrix}$.

We first show that the oscillator is not (standard) passive on $\mathcal X\times\mathbb R^2$. Choose any $\bar x\in\mathcal X$ with $1/2<\bar r:=\|\bar x\|<1$, and set $\bar u=-\mathcal J\bar x+\left(1-\frac{1}{\bar r}\right)\bar x$. Then $\dot x=f(\bar x,\bar u)=0$ and $\bar y=\bar x$. Consequently, every continuously differentiable state function $Q$ satisfies $\mathrm d_{\bar x}Q[f(\bar x,\bar u)]=0$.
However, since $\mathcal J^\top=-\mathcal J$,
$$\bar u^\top\bar y = \left(1-\frac{1}{\bar r}\right)\|\bar x\|^2 = \bar r(\bar r-1)<0.$$
The standard passivity inequality would therefore require $ 0= \mathrm d_{\bar x}Q[f(\bar x,\bar u)] \leq \bar u^\top\bar y<0$, which is impossible. Hence, no continuously differentiable nonnegative storage can establish standard passivity on the stated domain.

Now, we prove that the system is transversally passive. Choose the target input and output manifolds $\mathcal M_u=\{0\}$ and $\mathcal M_y=\mathbb S^1:=\{y\in\mathbb R^2:\|y\|=1\}$. Take the tubular neighborhoods $\mathcal N_u=\mathbb R^2$ and $\mathcal N_y= \{y\in\mathbb R^2:1/2<\|y\|<3/2\}$. The associated nearest-point retractions and transverse error maps are $r_u(u)=0$, $e_u(u)=u$, and $r_y(y)=\frac{y}{\|y\|}$, $e_y(y)=y-\frac{y}{\|y\|}$. The tubular domain is $\mathcal D= \{(x,u)\in\mathcal X\times\mathbb R^2: 1/2<\|x\|<3/2\}$. Choose the continuously differentiable nonnegative storage function $V(x)=\frac12(\|x\|-1)^2 =\frac12\|e_y(x)\|^2$. Its gradient satisfies $\nabla V(x)=e_y(x)$. Moreover, $e_y(x)$ is parallel to $x$, so $e_y(x)^\top\mathcal Jx=0$. Therefore, for every $(x,u)\in\mathcal D$,
\begin{equation*}
    \mathrm d_xV[f(x,u)] 
    =\langle e_u(u),e_y(y)\rangle-\|e_y(y)\|^2.
\end{equation*}
Thus, the system is output-strictly $(\{0\},\mathbb S^1)$-passive on $\mathcal D$; see Fig.~\ref{fig:oscillator}.
\end{example}

The results of Propositions \ref{prop:necessary_condition} and \ref{prop:sufficient_tpassivity}  provide basic necessary and sufficient tools for analyzing transversal passivity. We next explore how this notion relates to existing concepts in passivity theory.

\subsection{Connections to Existing Passivity Notions}
\label{ssec:vs_standard_passivity}
Existing passivity variants differ primarily in what their supply rate is referenced to --- a point, another trajectory, or an infinitesimal variation. Transversal passivity references it to a manifold. This recovers the first two notions as special cases and connects to the third at equilibria. Throughout this subsection we take $\varepsilon=\gamma=0$, the strict cases being analogous.

\emph{Equilibrium-referenced notions.} For affine targets $\mathcal M_u=u^\star+S_u$ and $\mathcal M_y=y^\star+S_y$, with $S_u,S_y\subseteq\mathbb R^m$ linear subspaces, the nearest-point projections are $r_u(u)=u^\star+\mathrm P_{S_u}(u-u^\star)$ and $r_y(y)=y^\star+\mathrm P_{S_y}(y-y^\star)$, so that
\begin{equation}\label{eq:affine_errors}
e_u(u)=\mathrm P_{S_u^\perp}(u-u^\star), \quad e_y(y)=\mathrm P_{S_y^\perp}(y-y^\star),
\end{equation}
where $\mathrm P_{S^\perp}:=I-\mathrm P_{S}$. Transversal passivity is therefore shifted dissipativity expressed in the normal coordinates. With $S_u=S_y=\{0\}$ it reduces to shifted passivity about $(u^\star,y^\star)$ \cite{monshizadeh2019shifted}, and with $u^\star=y^\star=0$ to standard passivity \eqref{eq:standard_passivity}. The same specialization applies to each equilibrium pair appearing in the EIP and MEIP inequalities \cite{hines2011EIP,burger2014duality}; those are families of singleton-target inequalities carrying a separate storage per equilibrium, and neither implies nor is implied by transversal passivity for a fixed pair of non-singleton targets.

\emph{Trajectory-referenced notions.} Incremental passivity references the supply rate to a second trajectory, and is exactly transversal passivity of the product system relative to the diagonals. Consider two solutions of \eqref{eq:dynsys} with states $(x_1,x_2)$, inputs $(u_1,u_2)$ and outputs $(y_1,y_2)$. The system is \emph{incrementally passive} if there exists a continuously differentiable $V_\Delta:\mathcal X\times\mathcal X\to\mathbb R_{\ge0}$ with
\begin{equation}\label{eq:incremental_passivity}
\dot V_\Delta(x_1,x_2)\le (u_1-u_2)^\top(y_1-y_2)
\end{equation}
along every pair of solutions.

\begin{proposition}\label{prop:incremental_diagonal}
Let $\mathcal U=\mathcal Y=\mathbb R^m$ and equip $\Sigma\times\Sigma$ with the diagonal targets $\Delta_u=\{(u,u):u\in\mathbb R^m\}$ and $\Delta_y=\{(y,y):y\in\mathbb R^m\}$. For a continuously differentiable nonnegative $V_\Delta$, the following are equivalent.
\begin{enumerate}
\item[(i)] $V_\Delta$ satisfies \eqref{eq:incremental_passivity} along every pair of solutions.
\item[(ii)] $V_\Delta/2$ is a transversal storage for $\Sigma\times\Sigma$ with respect to $(\Delta_u,\Delta_y)$ with indices $(\varepsilon,\gamma)=(0,0)$.
\end{enumerate}
\end{proposition}

\begin{proof}
The nearest-point projections onto the diagonals are the averages, so the transverse errors are the half-differences $e_{\Delta_u}=\tfrac12(u_1-u_2,\,u_2-u_1)$ and $e_{\Delta_y}=\tfrac12(y_1-y_2,\,y_2-y_1)$. Hence $\langle e_{\Delta_u},e_{\Delta_y}\rangle=\tfrac12(u_1-u_2)^\top(y_1-y_2)$, and the two dissipation inequalities coincide after scaling the storage by $2$.
\end{proof}

\emph{Infinitesimal notions.} Differential passivity references the supply rate to an infinitesimal variation and requires the inequality with supply $\delta u^\top\delta y$ along the variational system of every trajectory~\cite{van2013differentialpassivity}. Transversal passivity meets this notion only at equilibria: at singleton targets the projections in~\eqref{eq:necessary_TPassivity} are the identity, and Proposition~\ref{prop:necessary_condition} gives the differential passivity inequality of the linearization at $(x^\star,u^\star)$. This is an equilibrium-level consequence and does not establish differential passivity along all trajectories.

\section{Feedback Interconnection\\ and Compatibility Residuals}\label{sec:linear_feedback}
Much of the power of passivity theory is its compositional property. The negative feedback interconnection of two passive systems is again passive, with the sum of the subsystem storage functions as the closed-loop storage function. This property rests on an exact cancellation of the two supply rates: the interconnection is power conserving, so that $\langle u_1,y_1\rangle+\langle u_2,y_2\rangle=0$, and adding the subsystem inequalities $\dot Q_i\le\langle u_i,y_i\rangle$ leaves $\dot Q_1+\dot Q_2\le0$.

\begin{figure}[t]
\centering
\begin{tikzpicture}[
  >={Latex[length=2.2mm,width=1.6mm]},
  sys/.style={draw, thick, rounded corners=2pt, fill=blue!6,
              minimum width=13mm, minimum height=8.5mm},
  chan/.style={draw, thick, fill=gray!12,
              minimum width=10mm, minimum height=7mm},
  sig/.style={thick, ->, rounded corners=3pt},
  lbl/.style={font=\small, inner sep=2pt}
]
  \node[sys]  (S1) at (0, 0)      {$\Sigma_1$};
  \node[sys]  (S2) at (0,-2.0)    {$\Sigma_2$};
  \node[chan] (A)  at ( 2.3,-1.0) {$\Gamma$};
  \node[chan] (AT) at (-2.3,-1.0) {$-\Gamma^{\top}$};
  \draw[sig] (S1.east) -- node[lbl,above]{$y_1$} (A.north |- S1.east) -- (A.north);
  \draw[sig] (A.south) -- (A.south |- S2.east) -- node[lbl,below]{$u_2$} (S2.east);
  \draw[sig] (S2.west) -- node[lbl,below]{$y_2$} (AT.south |- S2.west) -- (AT.south);
  \draw[sig] (AT.north) -- (AT.north |- S1.west) -- node[lbl,above]{$u_1$} (S1.west);
  \node[font=\scriptsize, text=gray!70!black, anchor=west] at ($(A.east)+(0.05,0)$) {forward};
  \node[font=\scriptsize, text=gray!70!black, anchor=east] at ($(AT.west)+(-0.05,0)$) {reverse};
\end{tikzpicture}
\caption{Feedback interconnection of $\Sigma_1$ and $\Sigma_2$ through the forward channel $\Gamma$ and the reverse channel $-\Gamma^\top$.}
\label{fig:feedback_interconnection}
\end{figure}
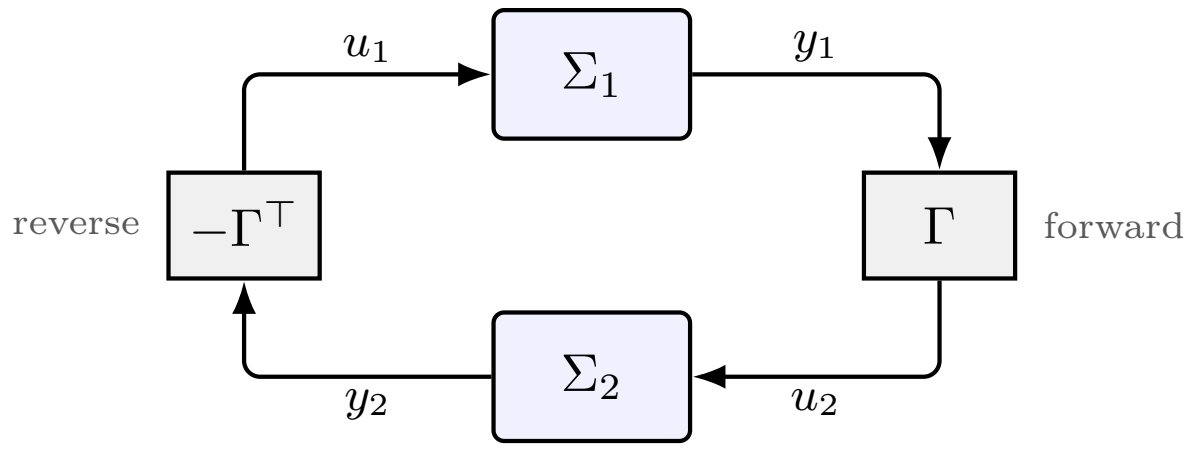
This power-conserving property also holds for the feedback interconnection in Fig.~\ref{fig:feedback_interconnection}, with inputs and outputs $u_i,y_i\in\mathbb R^{m_i}$ satisfying
\begin{equation}\label{eq:feedback_law}
u_1=-\Gamma^\top y_2, \quad u_2=\Gamma y_1, \quad \Gamma\in\mathbb R^{m_2\times m_1},
\end{equation}
where $\Gamma$ is the forward channel and $-\Gamma^\top$ the reverse channel. The port variables satisfying~\eqref{eq:feedback_law} form a separable Dirac structure~\cite[Proposition~2.5]{van2013pHgraph}, so that
$\langle u_1,y_1\rangle+\langle u_2,y_2\rangle=-\langle \Gamma^\top y_2,y_1\rangle+\langle \Gamma y_1,y_2\rangle=0$.

This cancellation is exactly what can fail for transversal passivity, because its supply rates are built from the transverse port errors rather than the ports themselves, while power conservation constrains only the ports. Consequently, transversal passivity does not compose in general. 

In this section, we first show that interconnecting two transversally passive systems through~\eqref{eq:feedback_law} leaves a non-cancellable residual supply in the closed-loop dissipation inequality. By analyzing the structure of the transversal supply rate and its residual, we then establish sufficient conditions under which transversal passivity composes, and show how closed-loop dissipation estimates can still be obtained when it does not.

\subsection{Closed-loop Realization and Dissipation}
Consider two nonlinear systems
\begin{equation}
\label{eq:sys_feedback}
\Sigma_i:
\left\{
\begin{aligned}
\dot x_i &= f_i(x_i,u_i),\\
y_i &= h_i(x_i,u_i),
\end{aligned}
\right.
\quad i=1,2,
\end{equation}
interconnected through \eqref{eq:feedback_law} with $\Gamma\in\mathbb R^{m_2\times m_1}$ as in Fig.~\ref{fig:feedback_interconnection}, where $x_i\in\mathcal X_i\subseteq\mathbb R^{n_i}$, $u_i\in\mathcal U_i\subseteq\mathbb R^{m_i}$ and $y_i\in\mathcal Y_i\subseteq\mathbb R^{m_i}$, with $\mathcal X_i,\mathcal U_i,\mathcal Y_i$ open, $h_i$ of class $C^1$ and $f_i$ locally Lipschitz. 

Adopting the notation of Section~\ref{sec:TP_def}, equip each subsystem $\Sigma_i$ with target manifolds $(\mathcal M_{u_i},\mathcal M_{y_i})$, tubular neighborhoods $(\mathcal N_{u_i},\mathcal N_{y_i})$, nearest-point retractions $r_{u_i}$ and $r_{y_i}$, and transverse port error maps $e_{u_i}$ and $e_{y_i}$. Assume that its transversal-passivity inequality holds with storage $V_i$ and indices $\varepsilon_i,\gamma_i\geq0$ on the nonempty domain
$$ 
\mathcal D_i:=\left\{(x_i,u_i)\in\mathcal X_i\times\mathcal U_i: u_i\in\mathcal N_{u_i},\; h_i(x_i,u_i)\in\mathcal N_{y_i}\right\}. 
$$

Both channels can be treated by the same construction. Let $B$ be a linear map from an output space carrying a target manifold $\mathcal M_y$, with tubular neighborhood $\mathcal N_y$ and transverse error map $e_y$, to an input space carrying $\mathcal M_u$, $\mathcal N_u$ and $e_u$. On the channel domain
\begin{equation}\label{eq:channel_domain}
\mathcal D_B:=\{y\in\mathcal N_y:\ By\in\mathcal N_u\},
\end{equation}
both error maps are well defined. 

\begin{figure}[t]
    \centering
    \includegraphics[width=\linewidth]{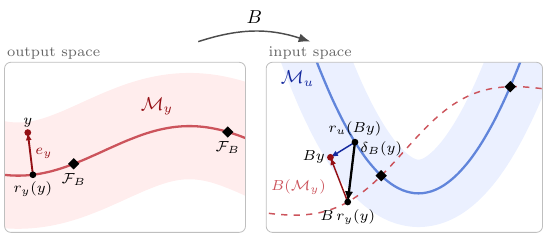}
    \caption{Compatibility defect for a linear channel $B$; the shaded bands are the tubes $\mathcal N_y$ and $\mathcal N_u$. The dashed curve is $B(\mathcal M_y)$, and the diamonds mark $\mathcal F_B$, discrete here, and its image. The defect $\delta_B(y)=B\,r_y(y)-r_u(By)$ (bold) is the mismatch between projecting then mapping and mapping then projecting.}
    \label{fig:defect}
\end{figure}

The channel transports transverse errors when $e_u(By)=B\,e_y(y)$. Since a linear channel need not commute with nonlinear nearest-point projections, this relation fails in general, and its failure is measured by the \emph{compatibility defect}, defined for $y\in\mathcal D_B$ by
\begin{equation}\label{eq:defect}
\delta_B(y):=e_u(By)-B\,e_y(y)=B\,r_y(y)-r_u(By).
\end{equation}
The channel also constrains which target points are attainable: the \emph{feasible target set}
\begin{equation}\label{eq:feasible_set}
\mathcal F_B:=\mathcal M_y\cap B^{-1}(\mathcal M_u)\subseteq\mathcal D_B
\end{equation}
collects the points of $\mathcal M_y$ that $B$ maps onto $\mathcal M_u$. Both objects are illustrated in Fig.~\ref{fig:defect}.

The forward and reverse channels of \eqref{eq:feedback_law} are the cases $B=\Gamma$ and $B=-\Gamma^\top$, respectively. The signal-space defect maps are $\delta_\Gamma$ and $\delta_{-\Gamma^\top}$. We denote the corresponding channel domains and feasible sets by $\mathcal D_{fc},\mathcal D_{rc},\mathcal F_{fc},\mathcal F_{rc}$, collected into the \emph{joint channel domain} $\mathcal D_\Gamma$ and the \emph{joint feasibility set} $\mathcal F_\Gamma$.
\begin{align}
\mathcal D_\Gamma&:=\mathcal D_{fc}\times\mathcal D_{rc},\label{eq:joint_channel_domain} \\ 
\mathcal F_\Gamma&:=\mathcal F_{fc}\times\mathcal F_{rc}, \label{eq:joint_feasible_target_set}
\end{align}
so that $\mathcal F_\Gamma\subseteq\mathcal D_\Gamma$.

For the closed-loop realization, since the output maps $h_i$ may contain direct feedthrough, algebraic well-posedness is not automatic. We therefore impose the following standing assumption.
\begin{assumption}[Algebraic well-posedness]\label{ass:linear_algebraic_well_posedness}
There exists an open set $\mathcal X_\Gamma\subseteq\mathcal X_1\times\mathcal X_2$ such that, for every $x=(x_1,x_2)\in\mathcal X_\Gamma$, the realized port maps
$$
\begin{aligned}
y_1(x)&=h_1(x_1,u_1(x)), & u_2(x)&=\Gamma y_1(x),\\
y_2(x)&=h_2(x_2,u_2(x)), & u_1(x)&=-\Gamma^\top y_2(x)
\end{aligned}
$$
have a unique solution $(y_1(x),u_2(x),y_2(x),u_1(x))$ in $\mathcal Y_1\times\mathcal U_2\times\mathcal Y_2\times\mathcal U_1$. 
\end{assumption}
We further assume that the realized port maps are $C^1$ on the algebraic well-posedness domain $\mathcal X_\Gamma$. Throughout the remainder of this work, we adopt Assumption~\ref{ass:linear_algebraic_well_posedness} and the $C^1$ regularity of the realized port maps as standing hypotheses and do not repeat them. The associated closed-loop maps, state-space sets, and preimages are understood relative to $\mathcal X_\Gamma$.

Within $\mathcal X_\Gamma$, we can define the realized output map and the closed-loop vector field by
\begin{equation}\label{eq:closed_loop_field}
h_\Gamma(x):=\bigl(y_1(x),y_2(x)\bigr),
\quad
F_{\mathrm{cl}}(x):=\begin{bmatrix}
f_1(x_1,u_1(x))\\
f_2(x_2,u_2(x))
\end{bmatrix},
\end{equation}
so that the closed loop is $\dot x=F_{\mathrm{cl}}(x)$.
Then $F_{\mathrm{cl}}$ is locally Lipschitz on $\mathcal X_\Gamma$. 

Restricting the realized ports to the subsystem tubular domains $\mathcal D_i$ gives the closed-loop tubular domain
\begin{equation}\label{eq:cl_certificate}
\mathcal D_{\mathrm{cl}} := \left\{x\in\mathcal X_\Gamma: (x_i,u_i(x))\in\mathcal D_i, \quad i=1,2 \right\},
\end{equation}
where the realized transverse port errors are defined and both transversal-passivity inequalities apply. We use the shorthand $e_{y_i}(x):=e_{y_i}(y_i(x))$ and $e_{u_i}(x):=e_{u_i}(u_i(x))$, for $i=1,2$. The realized defect maps are $\delta_{fc}(x):=\delta_\Gamma(y_1(x))$ and $\delta_{rc}(x):=\delta_{-\Gamma^\top}(y_2(x))$. From now on, channel-indexed symbols $\delta_B$ denote signal-space defect maps, whereas $\delta_{fc}$ and $\delta_{rc}$ denote their closed-loop realizations.  

Note that, by the feedback relations, $\mathcal D_{\mathrm{cl}}$ equals $h_\Gamma^{-1}(\mathcal D_\Gamma)$ and the closed-loop feasible state set is
\begin{equation}\label{eq:linear_feasible_state}
\begin{aligned}
\mathcal Z_{\mathrm{cl}}&:=h_\Gamma^{-1}(\mathcal F_\Gamma)\\
&=\{x\in\mathcal D_{\mathrm{cl}}:
e_{y_i}(x)=e_{u_i}(x)=0,\ i=1,2\}.
\end{aligned}
\end{equation}

The main point of this subsection is that the part of the supply that survives the interconnection can be attributed to the channels. In standard passivity, adding the subsystem inequalities cancels the supply rates because the interconnection is power conserving. Here the supply rates are built from transverse errors, and by~\eqref{eq:defect} the channels transport these errors only up to the defects, $e_{u_2}=\Gamma e_{y_1}+\delta_{fc}$ and $e_{u_1}=-\Gamma^\top e_{y_2}+\delta_{rc}$. The $\Gamma$-terms still cancel by adjointness, but the defect terms do not, and they remain as a residual supply.

To state this, define the composite storage $V_{\mathrm{cl}}(x):=V_1(x_1)+V_2(x_2)$ and the quadratic dissipation term 
$Q(x):=\sum_{i=1}^2\bigl(\varepsilon_i\|e_{y_i}(x)\|^2 +\gamma_i\|e_{u_i}(x)\|^2\bigr)$.

\begin{theorem}[Closed-loop dissipation and residual supply]\label{thm:linear_closed_loop_dissipation}
Consider subsystems $\Sigma_1$ and $\Sigma_2$ with storage functions $V_1$ and $V_2$. Then, for every $x\in\mathcal D_{\mathrm{cl}}$,
\begin{equation}\label{eq:closed_loop_mismatch}
\mathrm d_xV_{\mathrm{cl}}[F_{\mathrm{cl}}(x)] \leq s_\delta(x)-Q(x),
\end{equation}
where the \emph{residual transverse supply} $s_\delta$ is the total transverse supply of the two subsystems and admits the channel-defect form
\begin{equation}\label{eq:linear_residual_supply}
\begin{aligned}
s_\delta(x)&:=\textstyle\sum_{i=1}^2\langle e_{u_i}(x),e_{y_i}(x)\rangle\\
&\,=\langle\delta_{rc}(x),e_{y_1}(x)\rangle + \langle\delta_{fc}(x),e_{y_2}(x)\rangle.
\end{aligned}
\end{equation}
\end{theorem}

\begin{proof}
Summing the transversal-passivity inequalities of $\Sigma_1$ and $\Sigma_2$ on $\mathcal D_{\mathrm{cl}}$ gives~\eqref{eq:closed_loop_mismatch} with $s_\delta$ given by the first expression in~\eqref{eq:linear_residual_supply}. By~\eqref{eq:feedback_law} and~\eqref{eq:defect}, the realized transverse input errors decompose as $e_{u_2}(x)=\Gamma e_{y_1}(x)+\delta_{fc}(x)$ and $e_{u_1}(x)=-\Gamma^\top e_{y_2}(x)+\delta_{rc}(x)$. Substituting these relations into the first expression, the channel terms $\langle\Gamma e_{y_1},e_{y_2}\rangle-\langle\Gamma^\top e_{y_2},e_{y_1}\rangle$ cancel by adjointness, which leaves the second expression.
\end{proof}
The two expressions in~\eqref{eq:linear_residual_supply} indicate how the rest of this section proceeds. The first is the total transverse supply, which in standard passivity would vanish by power conservation. The second attributes the part that survives to the channels, and closed-loop dissipation estimates therefore reduce to analyzing these channel terms. Since each defect depends only on a channel and its prescribed manifolds, not on the subsystem dynamics, conditions under which the defects vanish can be checked in signal space, without the closed-loop realization; Subsection~\ref{ssec:vanish_residuals} develops such conditions. When the defects do not vanish, both factors in each term of the second expression still vanish on the closed-loop feasible state set, so the residual is locally quadratic there; Subsection~\ref{ssec:nonvanishing_residuals} examines this quadratic part.

\subsection{Vanishing Residuals and Exact Compatibility}\label{ssec:vanish_residuals}
If $s_\delta(x)=0$, transversal passivity composes directly and~\eqref{eq:closed_loop_mismatch} reduces to $\mathrm d_xV_{\mathrm{cl}}[F_{\mathrm{cl}}(x)]\leq-Q(x)$. By~\eqref{eq:linear_residual_supply}, this is achieved when both channel defects vanish, providing a geometric sufficient condition that can be checked directly in signal space without solving for the closed-loop realization. Exact compatibility is a property of a region of output signals, typically a neighborhood of part of $\mathcal M_y$, on which the defect is checked.

\begin{definition}[Exact channel compatibility]
\label{def:exact_compatibility}
The channel $B$ is \emph{exactly compatible} on a relatively open set $\Omega_B\subseteq\mathcal D_B$ if $\delta_B$ vanishes identically on $\Omega_B$. The interconnection is exactly compatible on $\Omega:=\Omega_{fc}\times\Omega_{rc}\subseteq\mathcal D_\Gamma$ if this holds for $B=\Gamma$ on $\Omega_{fc}$ and for $B=-\Gamma^\top$ on $\Omega_{rc}$.
\end{definition}
By~\eqref{eq:defect}, $\delta_B(y)=0$ is equivalent to $r_u(By)=B r_y(y)$, so exact compatibility means that the channel commutes with the prescribed nearest-point retractions on $\Omega_B$. In closed loop, this signal-space condition eliminates $s_\delta(x)$ and relates the realized transverse output errors to the feasible target $\mathcal F_\Gamma$.

\begin{proposition}[Closed-loop dissipation and feasible distance under exact compatibility]
\label{prop:exact_compatibility_target_preservation}
Suppose that the interconnection is exactly compatible on $\Omega\subseteq\mathcal D_\Gamma$. Then, for every $x\in h_\Gamma^{-1}(\Omega)$, the residual supply vanishes, $s_\delta(x)=0$, and the following two properties hold:
\begin{enumerate}
\item[(i)] \emph{Dissipation:} the closed loop dissipates at least at rate $Q$,
\begin{equation}\label{eq:linear_exact_dissipation}
\mathrm d_xV_{\mathrm{cl}}[F_{\mathrm{cl}}(x)]\leq-Q(x)\leq0;
\end{equation}
\item[(ii)] \emph{Feasible distance:} the transverse output errors measure the distance to the feasible target,
\begin{equation}\label{eq:linear_exact_distance_pullback}
\dist\bigl(h_\Gamma(x),\mathcal F_\Gamma\bigr)^2=\|e_{y_1}(x)\|^2+\|e_{y_2}(x)\|^2.
\end{equation}
\end{enumerate}
\end{proposition}
\begin{proof}
    Since $\Omega\subseteq\mathcal D_\Gamma$, we have $h_\Gamma^{-1}(\Omega)\subseteq h_\Gamma^{-1}(\mathcal D_\Gamma)=\mathcal D_{\mathrm{cl}}$. For every $x\in h_\Gamma^{-1}(\Omega)$, $h_\Gamma(x)\in\Omega$ implies that both realized defects vanish, so $s_\delta(x)=0$ and~\eqref{eq:linear_exact_dissipation} follows from~\eqref{eq:closed_loop_mismatch}. Moreover, for an exactly compatible channel $B$,~\eqref{eq:defect} gives $B r_y(y)=r_u(By)\in\mathcal M_u$ on its compatibility region, hence $r_y(y)\in\mathcal F_B$. Since $\mathcal F_B\subseteq\mathcal M_y$,
    \begin{align*}
    \dist(y,\mathcal M_y)\leq\dist(y,\mathcal F_B)\leq\|y-r_y(y)\|=\dist(y,\mathcal M_y).
    \end{align*}
    Applying this identity to the two realized outputs and using $\dist((z_1,z_2),A_1\times A_2)^2=\dist(z_1,A_1)^2+\dist(z_2,A_2)^2$ for the Euclidean product norm proves~\eqref{eq:linear_exact_distance_pullback}.
\end{proof}
In general, the realized defects $\delta_{fc}$ and $\delta_{rc}$ do not vanish, so the residual transverse supply $s_\delta(x)$ is nonzero and sign-indefinite, and~\eqref{eq:closed_loop_mismatch} does not automatically reduce to~\eqref{eq:linear_exact_dissipation}. Obtaining closed-loop dissipation estimates in this case requires analyzing the local structure of $s_\delta(x)$ around the feasible state set, which we do in the sequel.

\subsection{Local Structure of Nonvanishing Residuals}\label{ssec:nonvanishing_residuals}
Suppose that the four realized maps $e_{y_1},e_{y_2},\delta_{fc},\delta_{rc}$ are $C^2$ near a state $p\in\mathcal D_{\mathrm{cl}}$. Since $s_\delta(x)$ in~\eqref{eq:linear_residual_supply} is a sum of inner products of these maps, a second-order Taylor expansion about $p$ gives
\begin{align*}
    s_\delta(p+\xi)= &s_\delta(p)+\mathrm d_p s_\delta[\xi]+\frac12\Bigl(
    \langle\mathrm d_p^2\delta_{rc}[\xi,\xi],e_{y_1}(p)\rangle \\
    &+\langle\delta_{rc}(p),\mathrm d_p^2e_{y_1}[\xi,\xi]\rangle
    +\langle\mathrm d_p^2\delta_{fc}[\xi,\xi],e_{y_2}(p)\rangle \\
    &+\langle\delta_{fc}(p),\mathrm d_p^2e_{y_2}[\xi,\xi]\rangle\Bigr)+\xi^\top R_p\xi+o(\|\xi\|^2),
\end{align*}
where $R_p:=\operatorname{sym}\!\left((\mathrm d_p\delta_{rc})^\top\mathrm d_pe_{y_1}
+(\mathrm d_p\delta_{fc})^\top\mathrm d_pe_{y_2}\right)$, with $\operatorname{sym}(L):=(L+L^\top)/2$. By the first expression in~\eqref{eq:linear_residual_supply}, equivalently $\xi^\top R_p\xi=\sum_i\langle\mathrm d_pe_{u_i}[\xi],\mathrm d_pe_{y_i}[\xi]\rangle$ when $p\in\mathcal Z_{\mathrm{cl}}$, so $R_p$ is the quadratic form of the linearized total transverse supply. 

On the closed-loop feasible state set $\mathcal Z_{\mathrm{cl}}=h_\Gamma^{-1}(\mathcal F_\Gamma)$, all four realized maps $e_{y_1},e_{y_2},\delta_{fc},\delta_{rc}$ vanish by~\eqref{eq:defect} and~\eqref{eq:linear_feasible_state}. Consequently, when $p\in\mathcal Z_{\mathrm{cl}}$, all preceding terms in the expansion vanish and only 
$$s_\delta(p+\xi)=\xi^\top R_p\xi+o(\|\xi\|^2)$$ remains. One way to obtain closed-loop dissipation estimates near $\mathcal Z_{\mathrm{cl}}$ is to cancel this quadratic term so that $s_\delta(x)$ is locally cubic.

\begin{proposition}[Quadratic cancellation of the residual]\label{prop:second_order_cancellation}
Let $n:=n_1+n_2$. Suppose that the realized maps $e_{y_1},e_{y_2},\delta_{fc},\delta_{rc}$ are $C^2$ on an open set $\mathcal O\subseteq\mathcal D_{\mathrm{cl}}$, and let $\mathcal K\subseteq\mathcal Z_{\mathrm{cl}}\cap\mathcal O$ be nonempty and compact.

Then there exist constants $r>0$ and $C>0$ such that the open $r$-neighborhood of $\mathcal K$ is contained in $\mathcal O$ and the following two properties hold:
\begin{enumerate}
\item[(i)] \emph{Quadratic approximation:} for every $p\in\mathcal K$ and every $\xi\in\mathbb R^n$ with $\|\xi\|<r$, the residual supply is quadratic up to a cubic error,
\begin{equation}\label{eq:residual_quadratic_expansion}
\left|s_\delta(p+\xi)-\xi^\top R_p\xi\right|
\leq C\|\xi\|^3;
\end{equation}
\item[(ii)] \emph{Cubic residual under quadratic cancellation:} if $R_p=0$ for every $p\in\mathcal K$, then for every $x$ with $\dist(x,\mathcal K)<r$, we have $|s_\delta(x)|\leq C\dist(x,\mathcal K)^3$ and
\begin{equation}\label{eq:state_cubic_residual}
\mathrm d_xV_{\mathrm{cl}}[F_{\mathrm{cl}}(x)]
\leq -Q(x)+C\dist(x,\mathcal K)^3.
\end{equation}
\end{enumerate}
\end{proposition}

\begin{proof}
All four realized maps vanish on $\mathcal K\subseteq\mathcal Z_{\mathrm{cl}}$ by~\eqref{eq:defect} and~\eqref{eq:linear_feasible_state}. Since $\mathcal K$ is compact and $\mathcal O$ is open, we may choose $0<r\leq1$ such that $\mathcal N:=\{z\in\mathbb R^n:\dist(z,\mathcal K)\leq r\}$ is a compact subset of $\mathcal O$. For every $p\in\mathcal K$ and $\|\xi\|<r$, the segment $p+t\xi$, $t\in[0,1]$, lies in $\mathcal N$.

Since $\mathcal N$ is compact and the four maps are $C^2$, the constants
\begin{equation*}
L:=\max_{g}\,\sup_{z\in\mathcal N}\|\mathrm d_zg\|,\quad M:=\tfrac12\max_{g}\,\sup_{z\in\mathcal N}\|\mathrm d_z^2g\|,
\end{equation*}
with $g$ ranging over $\{e_{y_1},e_{y_2},\delta_{fc},\delta_{rc}\}$, are finite; we may take $L,M>0$. Since $g(p)=0$, Taylor's formula with integral remainder along the segment $p+t\xi$ gives $g(p+\xi)=\mathrm d_pg[\xi]+\rho_{g,p}(\xi)$ with
$\|\mathrm d_pg[\xi]\|\leq L\|\xi\|$ and $\|\rho_{g,p}(\xi)\|\leq M\|\xi\|^2$,
for all $p\in\mathcal K$ and $\|\xi\|<r$. Substituting these expansions into~\eqref{eq:linear_residual_supply}, the products of the first differentials sum to $\xi^\top R_p\xi$, and the remaining terms satisfy
\begin{align*}
\left|s_\delta(p+\xi)-\xi^\top R_p\xi\right|
&\leq 4LM\|\xi\|^3+2M^2\|\xi\|^4\\
&\leq (4LM+2M^2)\|\xi\|^3,
\end{align*}
because $\|\xi\|<r\leq1$. Thus~\eqref{eq:residual_quadratic_expansion} holds with $C:=4LM+2M^2$.

For each $p\in\mathcal K$, symmetry of $R_p$ implies that $R_p=0$ is equivalent to the directional cancellation condition
\begin{equation}\label{eq:second_order_cancellation}
\begin{aligned}
\langle\mathrm d_p\delta_{rc}[v],            \mathrm d_pe_{y_1}[v]\rangle + \langle\mathrm d_p\delta_{fc}[v], \mathrm d_pe_{y_2}[v]\rangle=0,
\end{aligned}
\end{equation}
for all $v\in\mathbb R^n$. When $R_p=0$ on $\mathcal K$, for any $x$ with $\dist(x,\mathcal K)<r$, compactness of $\mathcal K$ provides $p\in\mathcal K$ such that $\|x-p\|=\dist(x,\mathcal K)$. Taking $\xi=x-p$ in~\eqref{eq:residual_quadratic_expansion} gives $|s_\delta(x)|\leq C\dist(x,\mathcal K)^3$, and~\eqref{eq:closed_loop_mismatch} then yields~\eqref{eq:state_cubic_residual}.
\end{proof}
Checking $R_p=0$ directly requires the differentials of the realized maps, and hence the closed-loop realization. The structure of $R_p$ suggests a simpler sufficient condition: if both realized defects vanish to first order at $p$, that is, $\mathrm d_p\delta_{fc}=\mathrm d_p\delta_{rc}=0$, then $R_p=0$ regardless of the output errors. Since the realized defects are the signal-space defects evaluated along the realized outputs, this holds whenever the signal-space defects have vanishing differentials at the feasible targets.

\begin{definition}[First-order channel compatibility]\label{def:first_order_compatibility}
Suppose $\delta_B$ is of class $C^1$ on $\mathcal D_B$. The channel $B$ is \emph{first-order compatible along} $\mathcal S_B\subseteq\mathcal F_B$ if
\begin{equation}\label{eq:forward_first_order_compatibility}
\mathrm d_y\delta_B=0 \quad\text{for every } y\in\mathcal S_B.
\end{equation}
The interconnection is first-order compatible along $\mathcal S:=\mathcal S_{fc}\times\mathcal S_{rc}\subseteq\mathcal F_\Gamma$ if this holds for $B=\Gamma$ along $\mathcal S_{fc}$ and for $B=-\Gamma^\top$ along $\mathcal S_{rc}$.
\end{definition}
Differentiating~\eqref{eq:defect} shows that~\eqref{eq:forward_first_order_compatibility} is equivalent to
\begin{equation}\label{eq:first_order_forward_retraction}
\mathrm d_{By}r_u\circ B=B\circ\mathrm d_y r_y,
\quad y\in\mathcal S_B,
\end{equation}
which depends only on the channel and the target retractions. If the interconnection is first-order compatible along $\mathcal S$, then, for every $p\in h_\Gamma^{-1}(\mathcal S)$, the chain rule gives $\mathrm d_p\delta_{fc}=\mathrm d_{y_1(p)}\delta_\Gamma\circ\mathrm d_py_1=0$ and $\mathrm d_p\delta_{rc}=\mathrm d_{y_2(p)}\delta_{-\Gamma^\top}\circ\mathrm d_py_2=0$, so $R_p=0$. Under the hypotheses of Proposition~\ref{prop:second_order_cancellation}, first-order compatibility along $\mathcal S$ therefore yields~\eqref{eq:state_cubic_residual} for any compact $\mathcal K\subseteq h_\Gamma^{-1}(\mathcal S)\cap\mathcal O$.

Exact compatibility implies first-order compatibility. If the interconnection is exactly compatible on a relatively open $\Omega$, then $\delta_\Gamma$ and $\delta_{-\Gamma^\top}$ vanish identically on open sets, so their differentials vanish there as well, and the interconnection is first-order compatible along $\mathcal S=\Omega\cap(\mathcal M_{y_1}\times\mathcal M_{y_2})$. This set lies in $\mathcal F_\Gamma$ by the proof of Proposition~\ref{prop:exact_compatibility_target_preservation}. The converse fails when the channel matches the tangent spaces of the targets but not their curvature. In closed loop, $\Omega$ pulls back to an open set of states on which $s_\delta$ vanishes identically, whereas $\mathcal S$ pulls back to a subset $h_\Gamma^{-1}(\mathcal S)$ of the feasible state set $\mathcal Z_{\mathrm{cl}}$, near which $s_\delta$ is only cubically small.

When $R_p\neq0$, the quadratic term $\xi^\top R_p\xi$ does not vanish identically, and we retain its signed contribution. Under the regularity and compactness hypotheses of Proposition~\ref{prop:second_order_cancellation},~\eqref{eq:closed_loop_mismatch} and~\eqref{eq:residual_quadratic_expansion} give
\begin{equation}\label{eq:retained_quadratic_dissipation}
\mathrm d_{p+\xi}V_{\mathrm{cl}}[F_{\mathrm{cl}}(p+\xi)]\leq-Q(p+\xi)+\xi^\top R_p\xi+C\|\xi\|^3,
\end{equation}
for every $p\in\mathcal K$ and $\|\xi\|<r$. Obtaining a closed-loop dissipation estimate then requires comparing the available dissipation (such as $Q$ in~\eqref{eq:closed_loop_mismatch}) directly with the signed quadratic residual $\xi^\top R_p\xi$, which we do in Subsection~\ref{ssec:net_dissipation}.

In summary, the conditions of this section form a hierarchy. Exact compatibility on $\Omega$ eliminates the residual supply (Proposition~\ref{prop:exact_compatibility_target_preservation}). First-order compatibility along $\mathcal S$, or more generally $R_p=0$ on $\mathcal K$, makes it cubic (Proposition~\ref{prop:second_order_cancellation}(ii)). Otherwise, the quadratic residual $\xi^\top R_p\xi$ remains and must be compared with the available dissipation. Section~\ref{sec:linear_analysis} shows how each case yields port convergence. Table~\ref{tab:residual_hierarchy} summarizes the hierarchy and, for later reference, lists the corresponding convergence result of Section~\ref{sec:linear_analysis} for each case.

\begin{table}[t]
\centering
\caption{Compatibility conditions and residual supply. The last column lists the port-convergence result of Section~\ref{sec:linear_analysis} for each case and serves as a roadmap for that section.}
\label{tab:residual_hierarchy}
\footnotesize
\setlength{\tabcolsep}{3pt}
\renewcommand{\arraystretch}{1.25}
\begin{tabular}{@{}>{\raggedright\arraybackslash}p{2.5cm}>{\raggedright\arraybackslash}p{3.6cm}>{\raggedright\arraybackslash}p{1.3cm}@{}}
\toprule
Condition & Residual near $\mathcal Z_{\mathrm{cl}}$ & Result\\
\midrule
Exact on $\Omega$ ($\delta\equiv0$) & $s_\delta\equiv0$, $\dot V_{\mathrm{cl}}\leq-Q$ (Prop.~\ref{prop:exact_compatibility_target_preservation}) & Cor.~\ref{cor:exact_convergence}\\
$R_p=0$ on $\mathcal K$ (e.g., first-order along $\mathcal S$) & $|s_\delta|\leq C\dist(x,\mathcal K)^3$ (Prop.~\ref{prop:second_order_cancellation}(ii)) & Cor.~\ref{cor:cubic_residual_absorption}\\
$R_p\neq0$ & $\xi^\top R_p\xi+O(\|\xi\|^3)$, Eq.~\eqref{eq:retained_quadratic_dissipation} & Thm.~\ref{thm:net_dissipation_margin}\\
\bottomrule
\end{tabular}
\end{table}
\section{Port Convergence under\\ Transversal Passivity}\label{sec:linear_analysis}
This section establishes convergence of the realized port variables to their prescribed manifolds for the feedback interconnection of Section~\ref{sec:linear_feedback}. The closed-loop dissipation inequalities of Section~\ref{sec:linear_feedback} hold only on the domain where the transverse errors are defined and both transversal-passivity inequalities apply. A nonincreasing storage therefore keeps a trajectory in a storage sublevel set only while the trajectory remains in this domain, and the sublevel sets may reach its boundary or be unbounded. We first use additional dynamical information, such as conserved quantities, to confine trajectories to a compact subset of the domain, which ensures that they exist for all time. We then show that the ports converge to their prescribed manifolds under a dissipation margin, for each of the residual cases of Table~\ref{tab:residual_hierarchy}.

\subsection{Compact Confinement}\label{ssec:linear_invariance}
In classical Lyapunov arguments, a nonincreasing storage with compact sublevel sets keeps every trajectory in a compact set, so solutions exist for all time and remain bounded. Two features of the present setting prevent a direct use of this argument. First, the dissipation inequality holds only on a domain, so it constrains a trajectory only while the trajectory stays there. Second, the storage sublevel sets need not be compact. For example, they may be tubes about the target manifolds, which are often unbounded. We therefore complement the storage with a set that trajectories cannot leave, and require only that its intersection with a storage sublevel set be compact.

Throughout this section, let $\mathcal O\subseteq\mathcal D_{\mathrm{cl}}$, with $\mathcal D_{\mathrm{cl}}$ the closed-loop tubular domain~\eqref{eq:cl_certificate}, be a nonempty open analysis domain for the closed-loop dynamics $\dot x=F_{\mathrm{cl}}(x)$ in~\eqref{eq:closed_loop_field}. Since the dissipation inequalities apply only while a trajectory remains in $\mathcal O$, a subset of $\mathcal O$ is called \emph{forward invariant relative to} $\mathcal O$ if every solution initialized in it remains in it for as long as it remains in $\mathcal O$. The composite storage $V_{\mathrm{cl}}(x)=V_1(x_1)+V_2(x_2)$ of Theorem~\ref{thm:linear_closed_loop_dissipation} is continuously differentiable and nonnegative on $\mathcal O$, and for $c\geq0$ we write
$$
\mathcal L_c:=\{x\in\mathcal O:V_{\mathrm{cl}}(x)\leq c\}
$$
for its sublevel sets in $\mathcal O$.

A \emph{restriction} is a set $\mathcal H\subseteq\mathcal O$ that is forward invariant relative to $\mathcal O$. The trivial choice $\mathcal H=\mathcal O$ imposes no restriction. A conserved quantity provides a nontrivial one. Suppose that a continuously differentiable map $\chi:\mathcal O\to\mathbb R^k$ satisfies
\begin{equation}\label{eq:first_integral}
\mathrm d_x\chi[F_{\mathrm{cl}}(x)]=0
\end{equation}
for every $x\in \mathcal O$. Such a $\chi$ is called a \emph{first integral}, or \emph{conserved quantity}, of the closed loop. Each initial condition $x_0\in \mathcal O$ selects the conserved level set
\begin{equation}\label{eq:initial_leafwise}
\mathcal H
:=\left\{x\in \mathcal O:\chi(x)=\chi(x_0)\right\}.
\end{equation}
Whenever a solution remains in $\mathcal O$, the chain rule gives $\chi(x(t))=\chi(x_0)$, so $\mathcal H$ is a restriction. A conserved quantity constrains directions left unconstrained by $V_{\mathrm{cl}}$, without a further dissipation inequality in those directions. This use of conserved quantities to complement a storage function is in the same spirit as Casimir-based methods for interconnected Hamiltonian systems~\cite{maschke2000Casimir1}.

\begin{proposition}[Compact confinement]\label{prop:compact_confinement}
Let $\mathcal H\subseteq\mathcal O$ be a restriction, let $x_0\in\mathcal H$ and $c\geq V_{\mathrm{cl}}(x_0)$, and suppose that
\begin{enumerate}
\item[(i)] $\mathrm d_xV_{\mathrm{cl}}[F_{\mathrm{cl}}(x)]\leq0$ for every $x\in\mathcal H$, and
\item[(ii)] the closure of $\mathcal L_c\cap\mathcal H$ in $\mathbb R^{n}$, $n=n_1+n_2$, is a compact subset of $\mathcal O$.
\end{enumerate}
Then the solution initialized at $x_0$ is forward complete and remains in $\mathcal L_c\cap\mathcal H$ for all $t\geq0$. Moreover, for every continuous function $\psi:\mathcal O\to\mathbb R$, the map $t\mapsto\psi(x(t))$ is uniformly continuous on $[0,\infty)$.
\end{proposition}

\begin{proof}
Let $x:[0,t_{\max})\to\mathcal X_\Gamma$ be the maximal solution with $x(0)=x_0$, and write $\mathcal A:=\mathcal L_c\cap\mathcal H$. As long as $x$ remains in $\mathcal O$, relative forward invariance keeps it in $\mathcal H$, and (i) then gives $V_{\mathrm{cl}}(x(t))\leq V_{\mathrm{cl}}(x_0)\leq c$, so $x(t)\in\mathcal A$. If $\tau:=\sup\{t<t_{\max}:x([0,t])\subseteq\mathcal O\}<t_{\max}$, then $x([0,\tau))\subseteq\mathcal A$ and continuity give $x(\tau)\in\overline{\mathcal A}\subseteq\mathcal O$; since $\mathcal O$ is open, $x$ remains in $\mathcal O$ beyond $\tau$, a contradiction. Therefore $x(t)\in\mathcal A$ for all $t<t_{\max}$.

Next, suppose $t_{\max}<\infty$. Then $x([0,t_{\max}))$ lies in the compact set $\overline{\mathcal A}\subseteq\mathcal X_\Gamma$. Since $F_{\mathrm{cl}}$ in~\eqref{eq:closed_loop_field} is locally Lipschitz on $\mathcal X_\Gamma$, \cite[Theorem~3.3]{Khalil2008Nonlinear} extends the solution beyond $t_{\max}$, which is a contradiction. Thus, $t_{\max}=\infty$ and $x(t)\in\mathcal A$ for all $t\geq 0$.

It remains to show that $t\mapsto \psi(x(t))$ is uniformly continuous, which is what allows Barbalat's lemma to be applied in Theorem~\ref{thm:linear_port_convergence}. Continuity of $F_{\mathrm{cl}}$ on $\overline{\mathcal A}$ gives a constant $M_F\geq0$ such that $\|F_{\mathrm{cl}}(x)\|\leq M_F$ there. Thus, $\|x(t)-x(s)\|\leq M_F|t-s|$ for all $s,t\geq0$. Moreover, every continuous $\psi:\mathcal O\to\mathbb R$ is uniformly continuous on the compact set $\overline{\mathcal A}$, and therefore $t\mapsto\psi(x(t))$ is uniformly continuous on $[0,\infty)$.
\end{proof}

\begin{figure}[t]
    \centering
    \includegraphics[width=0.85\linewidth]{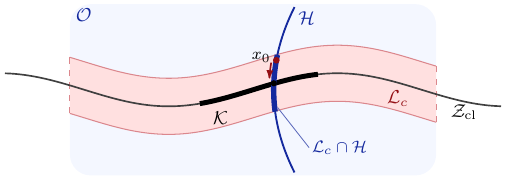}
    \caption{Sets of Section~\ref{sec:linear_analysis}. The sublevel set $\mathcal L_c$ (red) is a tube about $\mathcal Z_{\mathrm{cl}}$ that reaches the boundary of $\mathcal O$ (blue); a restriction $\mathcal H$ meets it in a compact set (thick) that confines the solution from $x_0$ to a neighborhood of $\mathcal K$ (bold), by Proposition~\ref{prop:compact_confinement}.}
    \label{fig:confinement_sets}
\end{figure}

With $\mathcal H=\mathcal O$, Proposition~\ref{prop:compact_confinement} reduces to the classical requirement that the closure of $\mathcal L_c$ be a compact subset of $\mathcal O$. This requirement does not follow from the closed-loop dissipation inequalities, which penalize only the realized transverse port errors: 
a bound on these errors need not bound the state, so $\mathcal L_c$ may reach the boundary of $\mathcal O$ or be unbounded. 
A restriction removes the need for compact sublevel sets, since only the intersection $\mathcal L_c\cap\mathcal H$ has to be compact. The restriction may depend on the initial condition, as the level set~\eqref{eq:initial_leafwise} does, since only the solution from $x_0$ must be confined. Fig.~\ref{fig:confinement_sets} illustrates the sets involved.


\begin{remark}[Conserved quantities from rank-deficient channels]\label{rem:channel_first_integrals}
Linear conserved quantities arise when a rank-deficient channel feeds integrating dynamics. Suppose that $\dot x_1=G_1u_1$ for a constant matrix $G_1$, and let $v$ satisfy $G_1^\top v\in\ker \Gamma$. Then $\tfrac{\mathrm d}{\mathrm dt}v^\top x_1=-(\Gamma G_1^\top v)^\top y_2=0$, so $\chi(x)=v^\top x_1$ satisfies~\eqref{eq:first_integral} for any output map $h_1$, and the level set~\eqref{eq:initial_leafwise} is an affine slice. Symmetrically, if $\dot x_2=G_2u_2$, then $G_2^\top v\in\ker \Gamma^\top$ makes $v^\top x_2$ conserved. Diffusively coupled networks posses this structure~\cite{arcak2007passivity,burger2014duality}. For integrator agents $\dot x_1=u_1$ and $\Gamma=E^\top$, with $E$ the incidence matrix of a connected graph, $\ker \Gamma=\operatorname{span}\{\mathbb 1\}$ and the conserved quantity is the network average, which selects the agreement point in average consensus.
\end{remark}

\begin{remark}[Other confinement mechanisms]\label{rem:linear_other_confinement}
Proposition~\ref{prop:compact_confinement} uses only that the restriction $\mathcal H$ contains $x_0$ and is forward invariant relative to $\mathcal O$. Besides conserved level sets, such a restriction may be obtained as a sublevel set of a continuous function that is nonincreasing along the closed loop. For example, along a gradient flow $\dot x=-\nabla\phi(x)$ of a convex function $\phi$, the distance to every minimizer is nonincreasing (Fej\'er monotonicity~\cite{bauschke2017convex}), so small closed discs centered at a minimizer near $x_0$ are compact restrictions even when the storage sublevel sets are unbounded. In each case, conditions (i) and (ii) of Proposition~\ref{prop:compact_confinement} must still be verified.
\end{remark}

Port convergence thus rests on two ingredients that are separate from the dissipation inequality itself: compact confinement of the closed-loop state, provided by Proposition~\ref{prop:compact_confinement}, and a dissipation margin that relates the dissipated power to the distance from the feasible target. This separation mirrors classical Lyapunov arguments, where boundedness and a decrease condition are established independently, and it allows each ingredient to be verified with tools suited to the system at hand. With forward completeness and uniform continuity ensured by compact confinement, we next establish port convergence under a dissipation margin.

\subsection{Port convergence under a Dissipation Margin}\label{ssec:exact_port_convergence}
Port convergence means that the realized inputs and outputs of $\Sigma_1$ and $\Sigma_2$ approach their prescribed manifolds:
\begin{equation}\label{eq:linear_port_convergence_objective}
\sum_{i=1}^{2}\Bigl(
\dist\bigl(u_i(x(t)),\mathcal M_{u_i}\bigr)^2
+\dist\bigl(y_i(x(t)),\mathcal M_{y_i}\bigr)^2
\Bigr)
\to 0.
\end{equation}

By the feedback law~\eqref{eq:feedback_law}, these inputs are determined by the realized outputs $h_\Gamma(x)=\bigl(y_1(x),y_2(x)\bigr)$, and all four ports lie on their prescribed manifolds precisely when $h_\Gamma(x)$ belongs to the joint feasibility set $\mathcal F_\Gamma=\mathcal F_{fc}\times\mathcal F_{rc}$ of~\eqref{eq:joint_feasible_target_set}. Since $\|-\Gamma^\top\|=\|\Gamma\|$, comparing $h_\Gamma(x)$ with any point of $\mathcal F_\Gamma$ gives
\begin{align*}
&\sum_{i=1}^{2}\bigl(\dist(u_i(x),\mathcal M_{u_i})^2+\dist(y_i(x),\mathcal M_{y_i})^2\bigr)\\
&\qquad\leq(1+\|\Gamma\|^2)\dist(h_\Gamma(x),\mathcal F_\Gamma)^2.
\end{align*}
Consequently, a sufficient condition for~\eqref{eq:linear_port_convergence_objective} is that the distance $\dist\bigl(h_\Gamma(x(t)),\mathcal F_\Gamma\bigr)$ approaches zero as $t\to\infty$.

The next theorem combines compact confinement with a dissipation margin that controls the distance to $\mathcal F_\Gamma$.
\begin{theorem}[Port convergence]\label{thm:linear_port_convergence}
Let $\mathcal H$, $x_0$, and $c$ satisfy the hypotheses of Proposition~\ref{prop:compact_confinement}. Suppose that there exist a continuous function $W:\mathcal O\to\mathbb R$ and a constant $\alpha_p>0$ such that the following two conditions hold for every $x\in\mathcal L_c\cap\mathcal H$:
\begin{enumerate}
\item[(i)] \emph{Dissipation:} the storage decreases at least at rate $W$,
\begin{equation}\label{eq:linear_general_dissipation_margin}
\mathrm d_xV_{\mathrm{cl}}[F_{\mathrm{cl}}(x)]\leq-W(x);
\end{equation}
\item[(ii)] \emph{Margin:} $W$ controls the distance to the feasible target,
\begin{equation}\label{eq:linear_general_port_coercivity}
W(x)\geq\alpha_p\dist\bigl(h_\Gamma(x),\mathcal F_\Gamma\bigr)^2.
\end{equation}
\end{enumerate}
Then the solution initialized at $x_0$ is forward complete and satisfies
\begin{equation}\label{eq:linear_general_output_convergence}
\lim_{t\to\infty}\dist\bigl(h_\Gamma(x(t)),\mathcal F_\Gamma\bigr)=0.
\end{equation}
\end{theorem}
\begin{proof}
By Proposition~\ref{prop:compact_confinement}, the solution is forward complete, remains in $\mathcal L_c\cap\mathcal H$, and $t\mapsto W(x(t))$ is uniformly continuous. Along this solution, \eqref{eq:linear_general_port_coercivity} gives $W(x(t))\geq 0$. Integrating~\eqref{eq:linear_general_dissipation_margin} over $[0,T]$ yields
$$
\int_0^T W(x(t))\,\mathrm dt \leq V_{\mathrm{cl}}(x_0)-V_{\mathrm{cl}}(x(T))\leq V_{\mathrm{cl}}(x_0).
$$
Thus, $W(x(\cdot))\in L^1([0,\infty))$, so Barbalat's lemma~\cite[Lemma~8.2]{Khalil2008Nonlinear} gives $\lim_{t\to\infty}W(x(t))=0$. The limit~\eqref{eq:linear_general_output_convergence} then follows from~\eqref{eq:linear_general_port_coercivity}.
\end{proof}
Theorem~\ref{thm:linear_port_convergence} applies whenever a dissipation margin~\eqref{eq:linear_general_port_coercivity} is available. When the residual supply vanishes, $W=Q$ is a natural choice, and output strictness of both subsystems supplies the margin.

\begin{corollary}[Port convergence under exact compatibility]\label{cor:exact_convergence}
Suppose that the interconnection is exactly compatible on $\Omega\subseteq\mathcal D_\Gamma$, that $\varepsilon_1>0$ and $\varepsilon_2>0$, and that $\mathcal O\subseteq h_\Gamma^{-1}(\Omega)$. Then the conclusions of Theorem~\ref{thm:linear_port_convergence} hold with $W=Q$ and $\alpha_p=\min\{\varepsilon_1,\varepsilon_2\}$ for every restriction $\mathcal H$, initial condition $x_0\in\mathcal H$, and $c\geq V_{\mathrm{cl}}(x_0)$ satisfying condition (ii) of Proposition~\ref{prop:compact_confinement}.
\end{corollary}
\begin{proof}
Proposition~\ref{prop:exact_compatibility_target_preservation}(i) gives~\eqref{eq:linear_general_dissipation_margin} with $W=Q$ on $\mathcal O$, which also yields condition (i) of Proposition~\ref{prop:compact_confinement}. Dropping the nonnegative input-strictness terms in $Q$ and using~\eqref{eq:linear_exact_distance_pullback} gives $Q(x)\geq\min\{\varepsilon_1,\varepsilon_2\}\bigl(\|e_{y_1}(x)\|^2+\|e_{y_2}(x)\|^2\bigr)=\min\{\varepsilon_1,\varepsilon_2\}\dist\bigl(h_\Gamma(x),\mathcal F_\Gamma\bigr)^2$, which is~\eqref{eq:linear_general_port_coercivity}.
\end{proof}

When the residual transverse supply $s_\delta(x)$ in~\eqref{eq:closed_loop_mismatch} does not vanish, applying Theorem~\ref{thm:linear_port_convergence} requires controlling this residual. By Subsection~\ref{ssec:nonvanishing_residuals}, for perturbations $\xi=x-p$ about a feasible state $p\in\mathcal Z_{\mathrm{cl}}$, the residual reduces locally to $s_\delta(p+\xi)=\xi^\top R_p\xi+o(\|\xi\|^2)$. The next two subsections address the cases where the quadratic residual vanishes ($R_p=0$) and where it remains nonzero ($R_p\neq0$), respectively.

\subsection{Port Convergence under Quadratic Cancellation}\label{ssec:quadratic_cancellation_convergence}
When $R_p=0$ on a compact feasible set, Proposition~\ref{prop:second_order_cancellation}(ii) bounds the residual supply by a cubic function of the distance to that set, and a quadratic dissipation margin absorbs it sufficiently close to the set. The argument uses the sets of Fig.~\ref{fig:confinement_sets}: a nonempty compact set $\mathcal K\subseteq\mathcal Z_{\mathrm{cl}}$ on which $R_p=0$, the open domain $\mathcal O$ of Proposition~\ref{prop:second_order_cancellation}, which contains $\mathcal K$, and a restriction $\mathcal H\subseteq\mathcal O$, on which alone the margin is required. Since the estimates are local, $\mathcal O$ may have to be shrunk to a smaller open neighborhood of $\mathcal K$. We then replace $\mathcal H$ by its intersection with the smaller set, which is again a restriction: a solution in this intersection stays in $\mathcal H$ as long as it stays in the smaller set.
\begin{corollary}[Cubic residual absorption]\label{cor:cubic_residual_absorption}
Suppose that the hypotheses of Proposition~\ref{prop:second_order_cancellation} hold with $R_p=0$ for every $p\in\mathcal K$. Let $\mathcal H\subseteq\mathcal O$ be a restriction, and suppose that $Q$ satisfies the quadratic margin
\begin{equation}\label{eq:quadratic_Q_margin}
Q(x)\geq\alpha\dist(x,\mathcal K)^2, \quad \forall x\in\mathcal H,
\end{equation}
for some $\alpha>0$. Then, after shrinking $\mathcal O$ to a sufficiently small open neighborhood of $\mathcal K$, the following statements hold.

\emph{(i) Net margin.} There exists a constant $L>0$ such that, for every $x\in\mathcal H$,
\begin{equation}\label{eq:cubic_net_margin}
\begin{aligned}
Q(x)-s_\delta(x)&\geq\tfrac{\alpha}{2}\dist(x,\mathcal K)^2\\
&\geq\tfrac{\alpha}{2L^2}\dist\bigl(h_\Gamma(x),\mathcal F_\Gamma\bigr)^2.
\end{aligned}
\end{equation}

\emph{(ii) Port convergence.} Let $x_0\in\mathcal H$ and $c\geq V_{\mathrm{cl}}(x_0)$ be such that the closure of $\mathcal L_c\cap\mathcal H$ is a compact subset of $\mathcal O$. Then the solution initialized at $x_0$ is forward complete and satisfies~\eqref{eq:linear_general_output_convergence}.
\end{corollary}
\begin{proof}
(i) Let $r$ and $C$ be the constants of Proposition~\ref{prop:second_order_cancellation}, so that $|s_\delta(x)|\leq C\dist(x,\mathcal K)^3$ whenever $\dist(x,\mathcal K)<r$, and the closed $r$-neighborhood $\mathcal N$ of $\mathcal K$ is a compact subset of $\mathcal O$. Let $L>0$ be an upper bound for $\|\mathrm d_xh_\Gamma\|$ on $\mathcal N$. For $\dist(x,\mathcal K)<r$, let $p\in\mathcal K$ be a nearest point to $x$; the segment from $p$ to $x$ lies in $\mathcal N$ and $h_\Gamma(p)\in\mathcal F_\Gamma$, so $\dist(h_\Gamma(x),\mathcal F_\Gamma)\leq\|h_\Gamma(x)-h_\Gamma(p)\|\leq L\dist(x,\mathcal K)$. Shrinking $\mathcal O$ to $\{x:\dist(x,\mathcal K)<\min\{r,\alpha/(2C)\}\}$ and combining these bounds with~\eqref{eq:quadratic_Q_margin} gives~\eqref{eq:cubic_net_margin}.

(ii) By~\eqref{eq:closed_loop_mismatch} and~\eqref{eq:cubic_net_margin}, the continuous function $W:=Q-s_\delta$ satisfies~\eqref{eq:linear_general_dissipation_margin}--\eqref{eq:linear_general_port_coercivity} on $\mathcal H$ with $\alpha_p=\alpha/(2L^2)$. In particular, $\mathrm d_xV_{\mathrm{cl}}[F_{\mathrm{cl}}(x)]\leq0$ on $\mathcal H$, so the hypotheses of Proposition~\ref{prop:compact_confinement} hold, and Theorem~\ref{thm:linear_port_convergence} yields the claim.
\end{proof}

The margin~\eqref{eq:quadratic_Q_margin} is the essential hypothesis of Corollary~\ref{cor:cubic_residual_absorption}. Since $Q$ penalizes only the realized transverse port errors, it may fail to detect certain state deviations from $\mathcal K$ at quadratic order. The following proposition identifies a geometric obstruction to obtaining such a margin from $Q$ on a full neighborhood.

\begin{proposition}[An obstruction to quadratic dissipation margins]\label{prop:quadratic-margin-obstruction}
Let $\mathcal K\subseteq\mathcal Z_{\mathrm{cl}}$ be a $C^2$ embedded submanifold in a neighborhood of $p\in\mathcal K$, and write $y_i^\star:=y_i(p)$ and $u_i^\star:=u_i(p)$ for $i=1,2$.

If there exists a state direction $v\in N_p\mathcal K\setminus\{0\}$ such that $\mathrm d_py_i[v]\in T_{y_i^\star}\mathcal M_{y_i}$ and $\mathrm d_pu_i[v]\in T_{u_i^\star}\mathcal M_{u_i}$ for $i=1,2$, then there is no constant $\alpha>0$ such that $Q(x)\geq\alpha\,\dist(x,\mathcal K)^2$ for every $x$ in a neighborhood of $p$.
\end{proposition}

\begin{proof}
For sufficiently small $\tau$, the state $x_\tau:=p+\tau v$ belongs to $\mathcal D_{\mathrm{cl}}$.
By Lemma~\ref{lem:tubular_projection_identities}(i), we have $\mathrm d_pe_{y_i}[v] = \mathrm P_{N_{y_i^\star}\mathcal M_{y_i}} \mathrm d_py_i[v] = 0$ and, likewise, $\mathrm d_pe_{u_i}[v]=0$ for $i=1,2$, where the differentials are taken along the realized port maps.

Since $p\in\mathcal K\subseteq\mathcal Z_{\mathrm{cl}}$, all four realized transverse port errors vanish at $p$. Their first-order expansions therefore imply $e_{y_i}(x_\tau)=o(|\tau|)$ and $e_{u_i}(x_\tau)=o(|\tau|)$ for $i=1,2$, so $Q(x_\tau)=o(\tau^2)$.

On the other hand, since $v\in N_p\mathcal K\setminus\{0\}$, the local nearest-point projection onto $\mathcal K$ yields
$$
\dist(x_\tau,\mathcal K)^2 = \tau^2\|v\|^2 + o(\tau^2),
$$
with $\|v\|>0$. Thus, $\displaystyle \lim_{\tau\to0,\ \tau\ne0} \frac{Q(x_\tau)}{\dist(x_\tau,\mathcal K)^2} = 0$, which excludes every positive quadratic lower bound on a neighborhood of $p$.
\end{proof}

If the realized transverse error maps are $C^2$, the same argument gives the sharper estimate $Q(x_\tau)=O(\tau^4)$. Proposition~\ref{prop:quadratic-margin-obstruction} concerns a full neighborhood of $p$. This is why Corollary~\ref{cor:cubic_residual_absorption} requires the margin only on the restriction $\mathcal H$: a restriction that excludes the obstructing directions $v$ can recover the margin, as in Subsection~\ref{ssec:fl_first_order}.

\subsection{Net Dissipation with Nonzero Quadratic Residuals}\label{ssec:net_dissipation}
When $R_p\neq0$, the quadratic residual $\xi^\top R_p\xi$ in~\eqref{eq:retained_quadratic_dissipation} has the same order as $Q(p+\xi)=\tfrac12\xi^\top\nabla^2Q(p)\xi+o(\|\xi\|^2)$. Unlike in Corollary~\ref{cor:cubic_residual_absorption}, shrinking the neighborhood absorbs only higher-order terms and cannot eliminate $\xi^\top R_p\xi$, so $\tfrac12\nabla^2Q(p)$ must be compared directly with $R_p$. The sets are those of Subsection~\ref{ssec:quadratic_cancellation_convergence}: a compact feasible set $\mathcal K$, a domain $\mathcal O$ that may be shrunk about $\mathcal K$, and a restriction $\mathcal H$. When $\mathcal H$ is a $C^2$ submanifold containing $\mathcal K$, the comparison can be carried out on its tangent space, in directions normal to $\mathcal K$.

\begin{theorem}[A restricted net-dissipation criterion]\label{thm:net_dissipation_margin}
Suppose that the realized maps $e_{y_1},e_{y_2},\delta_{fc},\delta_{rc}$ are $C^2$ on $\mathcal O$. Let $\mathcal H\subseteq\mathcal O$ be a restriction that is a $C^2$ embedded submanifold, and let $\mathcal K\subseteq\mathcal H\cap\mathcal Z_{\mathrm{cl}}$ be a nonempty compact $C^2$ embedded submanifold without boundary. For $p\in\mathcal K$, define the net quadratic form
\begin{equation}\label{eq:net_quadratic_form}
H_p:=\tfrac12\nabla^2Q(p)-R_p,
\end{equation}
and suppose that there exists $\lambda>0$ such that
\begin{equation}\label{eq:restricted_net_positivity}
v^\top H_pv\geq\lambda\|v\|^2, \quad  \forall p\in\mathcal K, \  v\in T_p\mathcal H\cap N_p\mathcal K.
\end{equation}
Then, after shrinking $\mathcal O$ to a sufficiently small open neighborhood of $\mathcal K$, the following statements hold.

\emph{(i) Net margin.} There exist constants $\kappa,L>0$ such that, for every $x\in\mathcal H$,
\begin{equation}\label{eq:net_dissipation_margin}
\begin{aligned}
Q(x)-s_\delta(x)&\geq\kappa\dist(x,\mathcal K)^2\\
&\geq\tfrac{\kappa}{L^2}\dist\bigl(h_\Gamma(x),\mathcal F_\Gamma\bigr)^2.
\end{aligned}
\end{equation}

\emph{(ii) Port convergence.} Let $x_0\in\mathcal H$ and $c\geq V_{\mathrm{cl}}(x_0)$ be such that the closure of $\mathcal L_c\cap\mathcal H$ is a compact subset of $\mathcal O$. Then the solution initialized at $x_0$ is forward complete and satisfies~\eqref{eq:linear_general_output_convergence}.

\emph{(iii) Local exponential convergence.} Suppose that $\mathcal K=\{p\}$ with $F_{\mathrm{cl}}(p)=0$, that $\mathcal H$ is relatively closed in $\mathcal O$, and that there exist $c_1,c_2>0$ such that
\begin{equation}\label{eq:leaf_storage_comparison}
c_1\|x-p\|^2\leq V_{\mathrm{cl}}(x)-V_{\mathrm{cl}}(p)\leq c_2\|x-p\|^2
\end{equation}
for all $x\in\mathcal H$ near $p$. Then, for every initial state $x(0)\in\mathcal H$ sufficiently close to $p$, the solution remains in $\mathcal O$, achieves port convergence, and satisfies
$$
\|x(t)-p\|\leq\sqrt{c_2/c_1}\,e^{-\kappa t/(2c_2)}\|x(0)-p\|,
\quad t\geq0.
$$
\end{theorem}
\begin{proof}
(i) By the feedback law, all $e_{y_i}$ and $e_{u_i}$ are $C^2$ on $\mathcal O$ and vanish on $\mathcal K\subseteq\mathcal Z_{\mathrm{cl}}$, so $Q$ is $C^2$ on $\mathcal O$ with $Q=0$ and $\mathrm d_pQ=0$ on $\mathcal K$. By Taylor's formula, Proposition~\ref{prop:second_order_cancellation}(i), and compactness of $\mathcal K$,
$$
Q(p+\xi)-s_\delta(p+\xi)=\xi^\top H_p\xi+o(\|\xi\|^2)
$$
uniformly for $p\in\mathcal K$. For $x\in\mathcal H$ near $\mathcal K$, let $p=r_{\mathcal K}(x)$ be its nearest point on $\mathcal K$ and $\xi=x-p$, so that $\xi\in N_p\mathcal K$. Since $\mathcal H$ is a $C^2$ embedded submanifold and $\mathcal K\subseteq\mathcal H$ is compact, finitely many slice charts of $\mathcal H$ cover $\mathcal K$, so near $\mathcal K$ the set $\mathcal H$ is the graph of a $C^2$ map over $T_p\mathcal H$ with constants uniform in $p\in\mathcal K$; embeddedness excludes other sheets of $\mathcal H$ accumulating on $\mathcal K$. Hence there is $C_{\mathcal H}>0$ such that $\eta:=\mathrm P_{T_p\mathcal H}\xi$ satisfies $\|\xi-\eta\|\leq C_{\mathcal H}\|\xi\|^2$. Since $T_p\mathcal K\subseteq T_p\mathcal H$, the projection $\mathrm P_{T_p\mathcal H}$ maps $N_p\mathcal K$ into itself, so $\eta\in T_p\mathcal H\cap N_p\mathcal K$ and~\eqref{eq:restricted_net_positivity} gives $\eta^\top H_p\eta\geq\lambda\|\eta\|^2$. As $H_p$ is bounded uniformly on $\mathcal K$, $\xi^\top H_p\xi=\eta^\top H_p\eta+O(\|\xi\|^3)$, and $\|\eta\|^2=\|\xi\|^2-\|\xi-\eta\|^2=\|\xi\|^2+O(\|\xi\|^4)$, so $Q(x)-s_\delta(x)\geq(\lambda+o(1))\|\xi\|^2$. Shrinking $\mathcal O$ gives the first inequality in~\eqref{eq:net_dissipation_margin} with $\kappa=\lambda/2$, and the Lipschitz bound for $h_\Gamma$ in the proof of Corollary~\ref{cor:cubic_residual_absorption}(i) gives the second.

(ii) As in the proof of Corollary~\ref{cor:cubic_residual_absorption}(ii),~\eqref{eq:closed_loop_mismatch} and~\eqref{eq:net_dissipation_margin} show that $W:=Q-s_\delta$ satisfies~\eqref{eq:linear_general_dissipation_margin}--\eqref{eq:linear_general_port_coercivity} on $\mathcal H$ and that $\mathrm d_xV_{\mathrm{cl}}[F_{\mathrm{cl}}(x)]\leq0$ there, so Proposition~\ref{prop:compact_confinement} and Theorem~\ref{thm:linear_port_convergence} yield the claim.

(iii) Let $B\subseteq\mathcal O$ be a closed ball of radius $\rho>0$ about $p$ on which~\eqref{eq:leaf_storage_comparison} holds for $x\in\mathcal H$, and let $B^\circ$ be its interior. By the left inequality in~\eqref{eq:leaf_storage_comparison}, $p$ is a strict minimizer of $V_{\mathrm{cl}}$ on $\mathcal H\cap B$. Fix $\varepsilon\in(0,c_1\rho^2)$ and let $\mathcal S:=\{x\in\mathcal H\cap B:V_{\mathrm{cl}}(x)\leq V_{\mathrm{cl}}(p)+\varepsilon\}$. Since $\mathcal H$ is relatively closed in $\mathcal O$ and $B\subseteq\mathcal O$ is compact, $\mathcal H\cap B$ and hence $\mathcal S$ are compact, and every $x\in\mathcal S$ satisfies $\|x-p\|\leq(\varepsilon/c_1)^{1/2}<\rho$, so $\mathcal S\subseteq B^\circ$. We apply Proposition~\ref{prop:compact_confinement} with $\mathcal O$, $\mathcal H$, and $c$ replaced by $B^\circ$, $\mathcal H\cap B^\circ$, and $V_{\mathrm{cl}}(p)+\varepsilon$. The set $\mathcal H\cap B^\circ$ is a restriction relative to $B^\circ$, because $\mathcal H$ is one relative to $\mathcal O\supseteq B^\circ$; condition (i) follows from~\eqref{eq:closed_loop_mismatch} and~\eqref{eq:net_dissipation_margin}; and condition (ii) holds because the relevant sublevel set is $\mathcal S$, which is compact and contained in $B^\circ$. If $x(0)\in\mathcal H$ and $\|x(0)-p\|\leq(\varepsilon/c_2)^{1/2}$, then $x(0)\in B^\circ$, since $c_1\leq c_2$, and $V_{\mathrm{cl}}(x(0))\leq V_{\mathrm{cl}}(p)+\varepsilon$ by the right inequality in~\eqref{eq:leaf_storage_comparison}. Hence the solution is forward complete and remains in $\mathcal S\subseteq B^\circ$. There,~\eqref{eq:closed_loop_mismatch},~\eqref{eq:net_dissipation_margin}, and~\eqref{eq:leaf_storage_comparison} give $\tfrac{\mathrm d}{\mathrm dt}\bigl(V_{\mathrm{cl}}(x)-V_{\mathrm{cl}}(p)\bigr)\leq-\kappa\|x-p\|^2\leq-(\kappa/c_2)\bigl(V_{\mathrm{cl}}(x)-V_{\mathrm{cl}}(p)\bigr)$; integrating and using~\eqref{eq:leaf_storage_comparison} gives the exponential bound, and~\eqref{eq:net_dissipation_margin} gives port convergence.
\end{proof}
Since $Q$ and $s_\delta$ vanish on $\mathcal Z_{\mathrm{cl}}$, the net margin~\eqref{eq:net_dissipation_margin} implies $\mathcal H\cap\mathcal Z_{\mathrm{cl}}\cap\mathcal O=\mathcal K$ for the shrunken $\mathcal O$. Thus condition~\eqref{eq:restricted_net_positivity} itself guarantees that $\mathcal K$ is isolated in $\mathcal H\cap\mathcal Z_{\mathrm{cl}}$, and it fails whenever $\mathcal H$ contains feasible states outside $\mathcal K$ arbitrarily close to it. As in Proposition~\ref{prop:quadratic-margin-obstruction}, the restriction must therefore exclude the feasible directions normal to $\mathcal K$.

Theorem~\ref{thm:net_dissipation_margin} also applies when $R_p=0$, in which case $H_p=\tfrac12\nabla^2Q(p)$ and~\eqref{eq:restricted_net_positivity} provides a second-order test for the restricted margin~\eqref{eq:quadratic_Q_margin} on $\mathcal H$. Since the transverse port errors vanish on $\mathcal K$, $\tfrac12v^\top\nabla^2Q(p)v=\sum_{i=1}^2\bigl(\varepsilon_i\|\mathrm d_pe_{y_i}[v]\|^2+\gamma_i\|\mathrm d_pe_{u_i}[v]\|^2\bigr)$, so the test requires that no direction in $T_p\mathcal H\cap N_p\mathcal K\setminus\{0\}$ be annihilated by the weighted port-error differentials. With $\mathcal H=\mathcal O$ and all strictness indices positive, this is exactly the negation of the hypothesis of Proposition~\ref{prop:quadratic-margin-obstruction} at every $p\in\mathcal K$, since $\mathrm d_pe_{y_i}[v]=0$ if and only if $\mathrm d_py_i[v]\in T_{y_i(p)}\mathcal M_{y_i}$, and likewise for the inputs. Under the regularity hypotheses of Theorem~\ref{thm:net_dissipation_margin}, the obstruction is then sharp: a quadratic margin on a full neighborhood of $\mathcal K$ exists if and only if no obstructing direction exists. Moreover, Corollary~\ref{cor:cubic_residual_absorption} and Theorem~\ref{thm:net_dissipation_margin} remain valid if $Q$ is replaced by any continuous (respectively, $C^2$) function $W_0:\mathcal O\to\mathbb R_{\geq0}$ satisfying
\begin{equation}\label{eq:available_dissipation}
\mathrm d_xV_{\mathrm{cl}}[F_{\mathrm{cl}}(x)]\leq s_\delta(x)-W_0(x),
\end{equation}
with $W_0=0$ on $\mathcal K$ in Theorem~\ref{thm:net_dissipation_margin}, for instance when an unrelaxed subsystem supply term is retained in place of a conservative quadratic bound $Q$.

In summary, applying the results of this section involves three steps. First, choose a restriction $\mathcal H$, for instance a conserved level set~\eqref{eq:initial_leafwise}, and verify compact confinement (Proposition~\ref{prop:compact_confinement}). Second, identify the residual case of Table~\ref{tab:residual_hierarchy}. Third, verify the corresponding margin: output strictness under exact compatibility (Corollary~\ref{cor:exact_convergence}), the quadratic margin~\eqref{eq:quadratic_Q_margin} on $\mathcal H$ when $R_p=0$ (Corollary~\ref{cor:cubic_residual_absorption}), or the net positivity condition~\eqref{eq:restricted_net_positivity} when $R_p\neq0$ (Theorem~\ref{thm:net_dissipation_margin}). Section~\ref{sec:fl} follows these steps.

\section{Case Study: Federated Learning With Heterogeneous Data}\label{sec:fl}
In federated learning, clients exchange model parameters or gradients with a server rather than raw data \cite{mcmahan2017fedavg,kairouz2021advances}. For overparameterized models the interpolating models can form a manifold \cite{liu2022loss}, and the central difficulty, data heterogeneity \cite{kairouz2021advances}, has a geometric meaning: each client has its own zero-loss manifold, and the clients agree only on their intersection. Throughout this section, the closed loop is fixed and only the server output target is varied. We show that heterogeneity precludes channel compatibility for every server target, yet port convergence can still be certified: via quadratic cancellation when the server target is tangent to the shared manifold, and via net dissipation on an invariant set when a nontangential target leaves a nonzero quadratic residual.

\subsection{Model, Transversal Passivity, and Compatibility}\label{ssec:fl_model}
\begin{figure}[!t]
\centering
\begin{tikzpicture}[
  >={Latex[length=2.2mm,width=1.6mm]},
  sys/.style={draw, thick, rounded corners=2pt, fill=blue!6, minimum height=7mm, align=center},
  cli/.style={draw, thick, rounded corners=2pt, fill=blue!6, minimum width=13mm, minimum height=6mm},
  chan/.style={draw, thick, fill=gray!12, minimum height=7mm, inner sep=3pt},
  sig/.style={thick, ->, rounded corners=3pt},
  wire/.style={thick, rounded corners=3pt},
  lbl/.style={font=\small, inner sep=2pt},
  tgt/.style={font=\scriptsize, text=gray!60!black, inner sep=1pt},
  dsg/.style={font=\scriptsize, text=red!70!black, inner sep=1pt}
]
  \node[sys] (S1) at (0,0) {$\Sigma_1$: server\\[-1pt] $\dot w=u_1$};
  \node[cli] (C1) at (0,-1.85) {$\nabla \ell_1$};
  \node      (Cd) at (0,-2.42) {$\vdots$};
  \node[cli] (CN) at (0,-3.05) {$\nabla \ell_N$};
  \node[draw, dashed, gray, rounded corners=3pt, inner sep=5pt, fit=(C1)(CN)] (S2) {};
  \node[font=\small, anchor=north] at (S2.south) {$\Sigma_2$: clients};
  \node[chan] (B) at ( 2.55,-0.95) {$\mathbb 1_N\!\otimes I_{1+d}$};
  \node[chan] (G) at (-2.55,-0.95) {$-\mathbb 1_N^{\top}\!\otimes I_{1+d}$};
  \draw[sig] (S1.east) -- node[lbl,above]{$y_1=w$} (B.north |- S1.east) -- (B.north);
  \node[dsg, anchor=south west] at ($(S1.east)+(0.15,-0.42)$) {$\mathcal M_{y_1}=\mathcal M$};
  \draw[wire] (B.south) -- (B.south |- CN.east);
  \draw[sig] (B.south |- C1.east) -- node[lbl,above]{$w$} (C1.east);
  \draw[sig] (B.south |- CN.east) -- node[lbl,above]{$w$} (CN.east);
  \fill (B.south |- C1.east) circle (1.3pt);
  \node[tgt, anchor=west] at ($(B.south |- CN.east)+(0.08,0.32)$) {$\mathcal M_{u_2}=\prod_i\mathcal Z_i$};
  \draw[wire] (C1.west) -- node[lbl,above]{$y_{2,1}$} (G.south |- C1.west);
  \draw[wire] (CN.west) -- node[lbl,above]{$y_{2,N}$} (G.south |- CN.west) -- (G.south |- C1.west);
  \draw[sig] (G.south |- C1.west) -- (G.south);
  \fill (G.south |- C1.west) circle (1.3pt);
  \node[tgt, anchor=east] at ($(G.south |- CN.west)+(-0.08,0.32)$) {$\mathcal M_{y_2}=\{0\}$};
  \draw[sig] (G.north) -- (G.north |- S1.west) -- node[lbl,above]{$u_1$} (S1.west);
  \node[tgt, anchor=north] at ($(G.north |- S1.west)!0.5!(S1.west)+(0,-0.12)$) {$\mathcal M_{u_1}=\{0\}$};
\end{tikzpicture}
\caption{Federated learning as the interconnection of Fig.~\ref{fig:feedback_interconnection}: the forward channel broadcasts the server model, and the reverse channel aggregates the client gradients. Port targets are in gray; only the server target $\mathcal M$ (red) is varied.}
\label{fig:fl_block}
\end{figure}
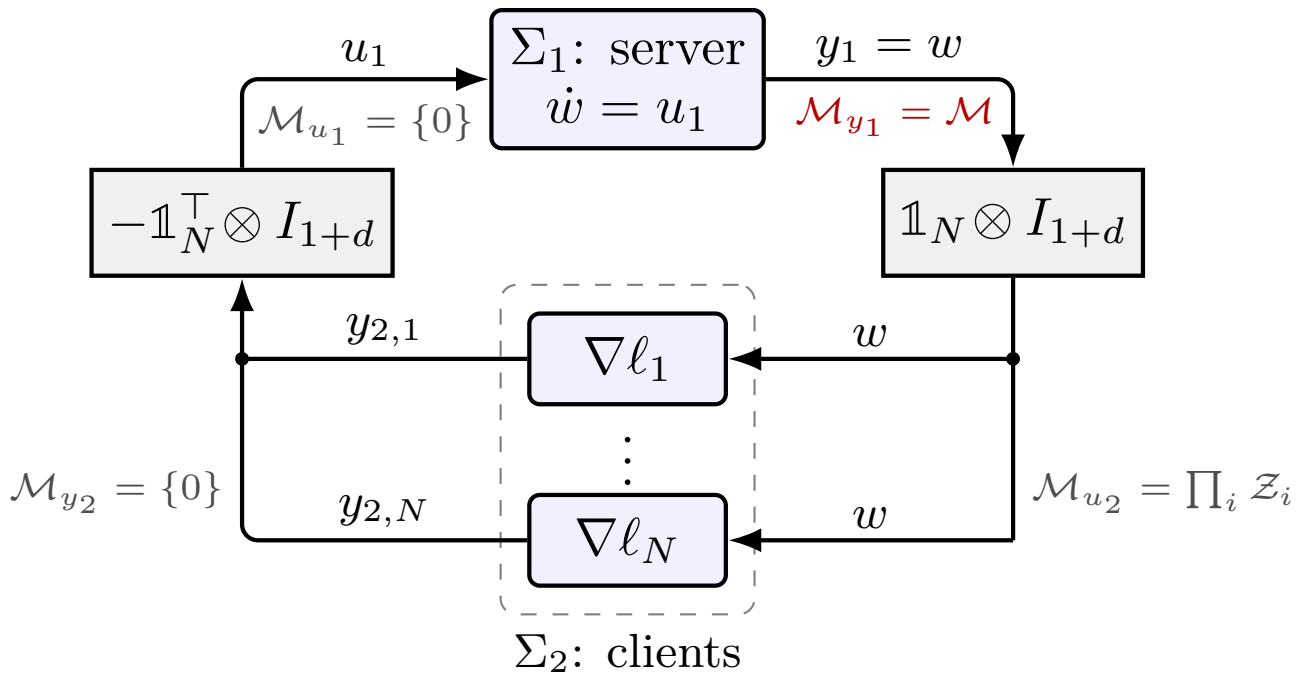

\emph{Learning problem.} Consider $N$ clients that jointly fit a linear predictor $\xi\mapsto\langle\theta,\xi\rangle$, $\theta\in\R^d$, through the factorization $\theta=ab$ with parameters $w=(a,b)\in\R\times\R^d$. This scalar--vector factorization is the simplest low-rank parametrization, and it makes the problem overparameterized, since each nonzero $\theta$ is represented by a curve of parameters $w$. Client $i$ holds one sample $(\xi_i,\tau_i)$ generated by a ground truth $\theta^\star$, with $\tau_i=\langle\theta^\star,\xi_i\rangle\neq0$, and has the squared loss
$$
\ell_i(w)=\tfrac12r_i(w)^2,\quad r_i(w):=a\langle b,\xi_i\rangle-\tau_i,
$$
with gradient $\nabla \ell_i=r_i\nabla r_i$ and $\nabla r_i=(\langle b,\xi_i\rangle,a\xi_i)$. We assume that the samples span $\R^d$, so that $X:=\sum_i\xi_i\xi_i^\top\succ0$, and write $m:=\|\theta^\star\|$ and $\hat\theta:=\theta^\star/m$.

\emph{Zero-loss manifolds and heterogeneity.} Client $i$ attains zero loss on $\mathcal Z_i:=\{r_i=0\}$. In the coordinates $s_i:=\langle b,\xi_i\rangle/\|\xi_i\|$ and $\mu_i:=\tau_i/\|\xi_i\|$, this is the hyperbolic cylinder $\{as_i=\mu_i\}$. The clients agree only on the common zero-loss set
$$
\mathcal Z:=\textstyle\bigcap_i\mathcal Z_i=\{ab=\theta^\star\},
$$
where the equality uses that the $\xi_i$ span $\R^d$. Since $ab=\theta^\star$ forces $b$ to be parallel to $\hat\theta$, this set lies in the plane
$$
\mathcal P:=\{(a,\beta\hat\theta):a,\beta\in\R\},
$$
where it is the hyperbola $a\beta=m$ with reach $\sqrt{2m}$. We use $(a,\beta)$, with $\beta=\langle b,\hat\theta\rangle$, as coordinates on $\mathcal P$. For $d=1$, all $\mathcal Z_i$ coincide with $\mathcal Z$. For $d\geq2$, they do not, since their normals $\nabla r_i$ point in different directions; this is the geometric form of data heterogeneity.

\emph{Training as a feedback interconnection.} In federated SGD with full participation~\cite{mcmahan2017fedavg}, taken in continuous time, the server broadcasts its model $w$, each client returns its gradient at $w$, and the server integrates the negative sum of these gradients. This is the interconnection of Fig.~\ref{fig:feedback_interconnection}, detailed in Fig.~\ref{fig:fl_block}, with
$$
\Sigma_1:\ \dot w=u_1,\ y_1=w,\quad \Sigma_2:\ y_{2,i}=\nabla \ell_i(u_{2,i}),
$$
and the broadcast channel $\Gamma=\mathbb 1_N\otimes I_{1+d}$. The server has no feedthrough and the clients are static, so the realized ports are explicit and Assumption~\ref{ass:linear_algebraic_well_posedness} holds. The closed loop is the gradient flow of the global loss $L:=\sum_i\ell_i$,
\begin{equation}\label{eq:fl_closed_loop}
\dot a=-\langle b,Xg\rangle,\quad \dot b=-aXg,\quad g:=ab-\theta^\star,
\end{equation}
and it conserves the imbalance $\chi:=a^2-\|b\|^2$ \cite{du2018balanced}, since $\dot\chi=-2a\langle b,Xg\rangle+2a\langle b,Xg\rangle=0$. The hyperbola $\mathcal Z$ has two branches, $a>0$ and $a<0$, related by the sign change $w\mapsto-w$. Along $\mathcal Z$ we have $b=\theta^\star/a$, so $\chi=a^2-m^2/a^2$ increases strictly with $|a|$ on each branch. Hence each imbalance value $\chi_0$ is attained at exactly one interpolant on each branch. We denote the one with $a>0$ by $p_{\chi_0}$, so that the other is $-p_{\chi_0}$. Since $\chi$ is conserved, $p_{\chi_0}$ is the only interpolant with $a>0$ that a trajectory starting at imbalance $\chi_0$ can reach. The balanced case $\chi_0=0$ gives $p_0=(\sqrt m,\theta^\star/\sqrt m)$, the interpolant of smallest norm, $\|p_0\|=\sqrt{2m}$.

\emph{Port targets.} The learning objective fixes $\mathcal M_{u_1}=\mathcal M_{y_2}=\{0\}$, so that the aggregated gradient vanishes, and $\mathcal M_{u_2}=\prod_i\mathcal Z_i$, so that each client receives a model it interpolates. The server target $\mathcal M_{y_1}=:\mathcal M$ is the design choice. Denoting the nearest-point projection and transverse error of a manifold $\mathcal M$ by $r_{\mathcal M}$ and $e_{\mathcal M}$, Lemma~\ref{lem:tubular_projection_identities}(ii) with $V_1=\tfrac12\dist(w,\mathcal M)^2$ gives $\dot V_1=\langle u_1,e_{\mathcal M}(w)\rangle$, so the server is $(\{0\},\mathcal M)$-passive for every $\mathcal M$.
Each client is a static map, so its transversal passivity reduces to a pointwise inequality between its transverse input error and its gradient. The reach of $\mathcal Z_i$ is $\sqrt{2|\mu_i|}$, the radius of curvature of the hyperbola $as_i=\mu_i$ at its vertex. We work on the tube of half this radius, $\eta_i:=\tfrac12\sqrt{2|\mu_i|}$, truncated to a ball of radius $R>0$,
\begin{equation}\label{eq:fl_client_tube}
\mathcal N_i:=\bigl\{u:\ \dist(u,\mathcal Z_i)<\eta_i,\ \|u\|<R\bigr\}.
\end{equation}
The truncation is needed because the strictness index below decreases with $R$.
\begin{lemma}[Client transversal passivity]\label{lem:fl_client_tp}
Let $R>0$. Then the following statements hold.
\begin{enumerate}
\item[(i)] \emph{Aiming:} for every client $i$ and every $u\in\mathcal N_i$,
\begin{equation*}
\langle e_{\mathcal Z_i}(u),\nabla \ell_i(u)\rangle\geq\frac{2}{3\|\xi_i\|^2R^2}\|\nabla \ell_i(u)\|^2.
\end{equation*}
\item[(ii)] \emph{Transversal passivity:} $\Sigma_2$ is output-strictly $(\prod_i\mathcal Z_i,\{0\})$-passive on $\prod_i\mathcal N_i$, with storage $V_2=0$ and index $\varepsilon_2=2/(3\max_i\|\xi_i\|^2R^2)$.
\end{enumerate}
\end{lemma}
\begin{proof}
Both sides of (i) vanish on $\mathcal Z_i$, so let $u\in\mathcal N_i\setminus\mathcal Z_i$. Since $\ell_i=\tfrac{\|\xi_i\|^2}{2}(as_i-\mu_i)^2$ depends on $u$ only through $z=(a,s_i)$, it suffices to treat the planar hyperbola $\{as=\mu_i\}$ with $\mu_i>0$. Write the transverse error of $z$ as $\nu n$, with base point $q$, unit normal $n=(q_2,q_1)/\|q\|$, and $|\nu|<\eta_i$. Expanding the quadratic $as-\mu_i$ about $q$ gives $\langle e_{\mathcal Z_i},\nabla \ell_i\rangle/\|\nabla \ell_i\|^2=(1+2\zeta)/\bigl(\|\xi_i\|^2(1+\zeta)\|z\|^2\bigr)$ with $\zeta:=\nu\mu_i/\|q\|^3$, and $|\zeta|<1/4$ because $\|q\|\geq\sqrt{2\mu_i}$. With $\|z\|<R$ this gives (i), and summing (i) over the clients gives (ii).
\end{proof}
The inequality of Lemma~\ref{lem:fl_client_tp} is the aiming condition of overparameterized learning \cite{liu2023aiming}, although the clients are neither passive nor monotone, since $\ell_i$ is nonconvex. A client aims at its own $\mathcal Z_i$, not at $\mathcal Z$: the supply $\langle e_{\mathcal Z},\nabla \ell_i\rangle$ can be negative, at fourth order, so $\mathcal M_{u_2}=\mathcal Z^N$ would not work.

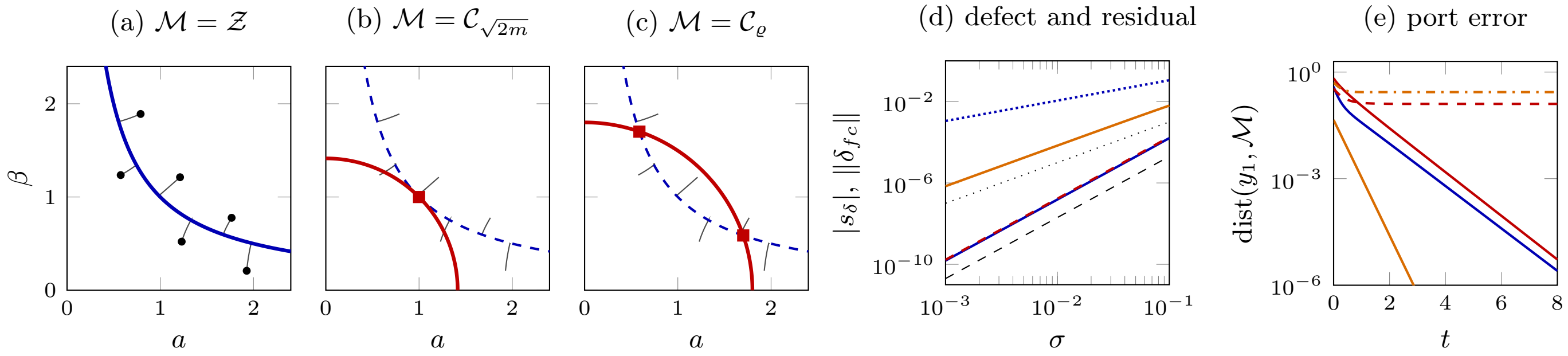
\begin{figure*}[t]
\centering
\pgfplotsset{flp/.style={scale only axis, width=0.142\textwidth, height=0.142\textwidth,
  tick label style={font=\scriptsize}, label style={font=\small}, title style={font=\small},
  ylabel shift=-3pt}}
\pgfplotsset{flphase/.style={flp, xmin=0, xmax=2.4, ymin=0, ymax=2.4, xlabel={$a$}, xtick={0,1,2}, ytick={0,1,2}}}
\begin{tikzpicture}
\begin{axis}[flphase, ylabel={$\beta$}, title={(a) $\mathcal M=\mathcal Z$}]
  \pgfplotsinvokeforeach{0,...,5}{\addplot[gray!60!black, thin] table[x=a,y=beta]{figures/fl/traj#1.dat};}
  \addplot[blue!70!black, very thick] table[x=a,y=beta]{figures/fl/Z.dat};
  \addplot[only marks, mark=*, mark size=1.1pt, black] table[x=a,y=beta]{figures/fl/ics.dat};
\end{axis}
\end{tikzpicture}\hfill
\begin{tikzpicture}
\begin{axis}[flphase, yticklabels={}, title={(b) $\mathcal M=\mathcal C_{\sqrt{2m}}$}]
  \pgfplotsinvokeforeach{0,...,5}{\addplot[gray!60!black, thin] table[x=a,y=beta]{figures/fl/traj#1.dat};}
  \addplot[blue!70!black, thick, dashed] table[x=a,y=beta]{figures/fl/Z.dat};
  \addplot[red!75!black, very thick] table[x=a,y=beta]{figures/fl/CF.dat};
  \addplot[only marks, mark=square*, mark size=1.8pt, red!75!black] table[x=a,y=beta]{figures/fl/FF.dat};
\end{axis}
\end{tikzpicture}\hfill
\begin{tikzpicture}
\begin{axis}[flphase, yticklabels={}, title={(c) $\mathcal M=\mathcal C_{\varrho}$}]
  \pgfplotsinvokeforeach{0,...,5}{\addplot[gray!60!black, thin] table[x=a,y=beta]{figures/fl/traj#1.dat};}
  \addplot[blue!70!black, thick, dashed] table[x=a,y=beta]{figures/fl/Z.dat};
  \addplot[red!75!black, very thick] table[x=a,y=beta]{figures/fl/CI.dat};
  \addplot[only marks, mark=square*, mark size=1.8pt, red!75!black] table[x=a,y=beta]{figures/fl/FI.dat};
\end{axis}
\end{tikzpicture}\hfill
\begin{tikzpicture}
\begin{loglogaxis}[flp, xmin=1e-3, xmax=1e-1, ymin=1e-11, ymax=1,
  xtick={1e-3,1e-2,1e-1}, ytick={1e-10,1e-6,1e-2},
  xlabel={$\sigma$}, ylabel={$|s_\delta|$, $\|\delta_{fc}\|$}, title={(d) defect and residual}]
  \addplot[blue!70!black, thick, densely dotted] table[x=d,y=defE]{figures/fl/residual.dat};
  \addplot[blue!70!black, thick] table[x=d,y=E]{figures/fl/residual.dat};
  \addplot[red!75!black, thick, dashed] table[x=d,y=F]{figures/fl/residual.dat};
  \addplot[orange!85!black, thick] table[x=d,y=I]{figures/fl/residual.dat};
  \addplot[black, thin, dashed, domain=1e-3:1e-1] {0.02*x^3};
  \addplot[black, thin, dotted, domain=1e-3:1e-1] {0.1*x^2};
\end{loglogaxis}
\end{tikzpicture}\hfill
\begin{tikzpicture}
\begin{semilogyaxis}[flp, xmin=0, xmax=8, ymin=1e-6, ymax=2, ytick={1e-6,1e-3,1},
  xlabel={$t$}, ylabel={$\dist(y_1,\mathcal M)$}, title={(e) port error}]
  \addplot[blue!70!black, thick] table[x=t,y=E]{figures/fl/porterr.dat};
  \addplot[red!75!black, thick] table[x=t,y=Fbal]{figures/fl/porterr.dat};
  \addplot[red!75!black, thick, dashed] table[x=t,y=Funb]{figures/fl/porterr.dat};
  \addplot[orange!85!black, thick] table[x=t,y=err]{figures/fl/porterr_matched.dat};
  \addplot[orange!85!black, thick, dashdotted] table[x=t,y=I]{figures/fl/porterr.dat};
\end{semilogyaxis}
\end{tikzpicture}
\caption{Heterogeneous data with $d=2$, $\theta^\star=(0.8,0.6)$, $\xi_1=(1,0)$, $\xi_2=(0.5,1)$, $\xi_3=(1,-0.5)$, $\varrho=1.8$. (a)--(c) The same trajectories (gray) projected onto $\mathcal P$, with $\mathcal Z$ (blue), the circle targets (red), and their feasible points (squares). (d) Defect $\|\delta_{fc}\|$ for $\mathcal M=\mathcal Z$ (dotted) and residual $|s_\delta|$ for the three targets (blue, red dashed, orange) along perturbations of size $\sigma$ from a feasible point; reference slopes $\sigma^2$ and $\sigma^3$. (e) Server port error for $\mathcal Z$ (blue), the minimum circle with $\chi_0=0$ (red) and $\chi_0\neq0$ (red dashed), and the larger circle on matched (orange) and unmatched (orange dash-dotted) imbalance levels. The matched trajectory starts at $b(0)=(a_p^{-1}+0.03)\hat\theta+0.04(-0.6,0.8)$ on $\{\chi=\chi(p)\}$; integration by fourth-order Runge--Kutta with step $0.002$.}
\label{fig:fl_phase}
\end{figure*}

Since $\delta_{rc}\equiv0$ and $e_{u_2}(\Gamma w)=(e_{\mathcal Z_i}(w))_i$, the forward defect, the residual supply, and the feasible sets are
\begin{equation}\label{eq:fl_defect}
\begin{aligned}
&\delta_{fc}(w)=\bigl(r_{\mathcal M}(w)-r_{\mathcal Z_i}(w)\bigr)_i,\\
&s_\delta(w)=\textstyle\sum_i\langle\delta_{fc,i}(w),\nabla \ell_i(w)\rangle,\\
&\mathcal F_\Gamma=(\mathcal M\cap\mathcal Z)\times\{0\},\quad \mathcal Z_{\mathrm{cl}}=\mathcal M\cap\mathcal Z,
\end{aligned}
\end{equation}
and Theorem~\ref{thm:linear_closed_loop_dissipation} gives $\dot V_{\mathrm{cl}}\leq s_\delta-Q$ with $V_{\mathrm{cl}}=V_1$ and $Q=\varepsilon_2\sum_i\|\nabla \ell_i\|^2$.
\begin{proposition}[Heterogeneity and compatibility]\label{prop:fl_heterogeneity}
Let $\mathcal M$ be a server target and let $p=(a_p,b_p)\in\mathcal M\cap\mathcal Z$ be a feasible point. Then the following statements hold.
\begin{enumerate}
\item[(i)] \emph{First-order compatibility:} the interconnection is first-order compatible along $\{p\}\times\{0\}$ if and only if $T_p\mathcal M=T_p\mathcal Z_i$ for every client $i$. For $d\geq2$, this condition fails for every server target.
\item[(ii)] \emph{Quadratic cancellation:} $R_p=0$ if and only if $T_p\mathcal M\subseteq T_p\mathcal Z$.
\end{enumerate}
\end{proposition}
\begin{proof}
(i) Lemma~\ref{lem:tubular_projection_identities}(i) gives $\mathrm d_p\delta_{fc,i}=\mathrm P_{T_p\mathcal M}-\mathrm P_{T_p\mathcal Z_i}$, which vanishes if and only if $T_p\mathcal M=T_p\mathcal Z_i$. Since each $\mathcal Z_i$ is a hypersurface, this equality for every $i$ forces the normals $\nabla r_i(p)=(\langle b_p,\xi_i\rangle,a_p\xi_i)$ to be pairwise parallel. As $a_p\neq0$ on $\mathcal Z$, the $\xi_i$ are then pairwise parallel and span at most a line, which contradicts $X\succ0$ when $d\geq2$. (ii) Since $r_i(p)=0$, $\mathrm d_p\nabla \ell_i[v]=\langle\nabla r_i(p),v\rangle\nabla r_i(p)$, which is normal to $\mathcal Z_i$. In~\eqref{eq:second_order_cancellation}, $v^\top R_pv$ therefore reduces to $v^\top\mathrm P_{T_p\mathcal M}Gv$ with $G:=\sum_i\nabla r_i(p)\nabla r_i(p)^\top\succeq0$ and $\ker G=\ker\mathrm d_p(ab)=T_p\mathcal Z$. It vanishes for all $v$ iff $T_p\mathcal M\subseteq\ker G$.
\end{proof}
In words, first-order compatibility would require the server target to be tangent to every client's zero-loss manifold at once, which heterogeneity rules out, whereas quadratic cancellation only requires it to be tangent to their intersection $\mathcal Z$. We compare three server targets, each encoding a different specification for the trained server model. The target $\mathcal M=\mathcal Z$ asks only that the server model interpolate the data of all clients, which is the federated learning objective itself. The circles $\mathcal C_\varrho:=\{w\in\mathcal P:\|w\|=\varrho\}$ in addition prescribe the norm of the trained parameters, a simple way to express a preference among the many interpolants of an overparameterized model. The radius $\varrho=\sqrt{2m}$ selects the minimum-norm interpolant $p_0$, while a radius $\varrho>\sqrt{2m}$ prescribes a larger norm, attained by two interpolants on each branch of $\mathcal Z$. Geometrically, the three targets meet $\mathcal Z$ along all of $\mathcal Z$, tangentially at a single point, and transversally, respectively. By Proposition~\ref{prop:fl_heterogeneity}(ii), this geometry determines the residual: $R_p=0$ for the first two targets and $R_p\neq0$ for the third; see Fig.~\ref{fig:fl_phase}. The first two instantiate the row $R_p=0$ of Table~\ref{tab:residual_hierarchy} (Corollary~\ref{cor:cubic_residual_absorption}) and the third the row $R_p\neq0$ (Theorem~\ref{thm:net_dissipation_margin}). The row of exact compatibility is absent, since exact compatibility implies first-order compatibility, which heterogeneity rules out for $d\geq2$.

\subsection{Shared Target: Quadratic Cancellation and Implicit Bias}\label{ssec:fl_exact}
Let $\mathcal M=\mathcal Z$. The defect grows linearly (Fig.~\ref{fig:fl_phase}(d)), yet by Proposition~\ref{prop:fl_heterogeneity}(ii) we have $R_p=0$ along all of $\mathcal Z$: the linear part of each client's defect is tangent to $\mathcal Z_i$, while its output varies normally to $\mathcal Z_i$. Work on the tube $\mathcal O:=\{w:\dist(w,\mathcal Z)<\eta,\ \|w\|<R\}$ with $\eta<\min\{\sqrt{2m},\min_i\eta_i\}$, where all subsystem inequalities hold because $\dist(w,\mathcal Z_i)\leq\dist(w,\mathcal Z)$.

Because $\mathcal Z$ is unbounded, the storage sublevel sets $\mathcal L_c$ reach the boundary of $\mathcal O$; the conserved imbalance confines them through the restriction $\mathcal H=\{\chi=\chi_0\}$ (Proposition~\ref{prop:compact_confinement}).
With $p=r_{\mathcal Z}(w)$ and $e=w-p$, $g=\mathrm d_p(ab)[e]+e_ae_b$ and $\|\mathrm d_p(ab)\|=\|p\|$ give
\begin{equation*}
\|w\|^4=\chi_0^2+4\|ab\|^2\leq\chi_0^2+4\bigl(m+\eta\|w\|+\tfrac32\eta^2\bigr)^2.
\end{equation*}
so for $R>\eta+\sqrt{|\chi_0|+2m+4\eta^2}$ and $c<\eta^2/2$, the closure of $\mathcal L_c\cap\{\chi=\chi_0\}$ is a compact subset of $\mathcal O$, on which $|a|$ and the first coordinate of $p$ exceed some $a_->0$ in magnitude.Since $\mathrm d_p(ab)$ has smallest singular value at least $a_-$ on $N_p\mathcal Z$ and $\|\nabla r_i\|\geq|a|\|\xi_i\|$, we obtain
\begin{equation}\label{eq:fl_gradient_bound}
\textstyle\sum_i\|\nabla \ell_i(w)\|^2\geq c_{\mathcal Z}\|e\|^2
\end{equation}
for $\|e\|\leq a_-$, with $c_{\mathcal Z}:=\tfrac14\lambda_{\min}(X)\min_i\|\xi_i\|^2a_-^4$. Corollary~\ref{cor:cubic_residual_absorption} now applies on the positive branch with $\mathcal K=\{p_{\chi_0}\}=\mathcal Z_{\mathrm{cl}}\cap\mathcal H$, where $R_{p_{\chi_0}}=0$. Since $\chi$ is strictly monotone along each branch of $\mathcal Z$, the level set $\mathcal H$ is transverse to $\mathcal Z$ at $p_{\chi_0}$, so $\|w-p_{\chi_0}\|\leq C_T\|e\|$ for $w\in\mathcal H$ near $p_{\chi_0}$ and some $C_T>0$. Together with~\eqref{eq:fl_gradient_bound}, this gives the margin~\eqref{eq:quadratic_Q_margin}, $Q(w)\geq\varepsilon_2c_{\mathcal Z}\|e\|^2\geq(\varepsilon_2c_{\mathcal Z}/C_T^2)\|w-p_{\chi_0}\|^2$, on $\mathcal H$ near $p_{\chi_0}$. Hence, for $c$ small enough that the closure of $\mathcal L_c\cap\mathcal H$ lies in the neighborhood of Corollary~\ref{cor:cubic_residual_absorption}, the ports converge. Since the solution remains in $\mathcal H$ near $p_{\chi_0}$ and $e(t)\to0$, its $\omega$-limit set lies in $\mathcal Z\cap\mathcal H=\{p_{\chi_0}\}$ there, so $w(t)\to p_{\chi_0}$, and symmetrically $w(t)\to-p_{\chi_0}$ on the negative branch. Heterogeneity leaves the local implicit bias of factorized models unchanged, since balanced initializations ($\chi_0=0$) converge to $p_0$. What changes is the certificate: quadratic cancellation replaces exact compatibility, which holds for homogeneous data.

\subsection{Minimum-Norm Target: The Margin Obstruction}\label{ssec:fl_first_order}
Let $\mathcal M=\mathcal C_{\sqrt{2m}}$, so $\mathcal Z_{\mathrm{cl}}=\mathcal M\cap\mathcal Z=\{p_0\}$ on the positive branch, with $T_{p_0}\mathcal M=T_{p_0}\mathcal Z$. Taking $\mathcal K=\{p_0\}$, Proposition~\ref{prop:fl_heterogeneity}(ii) gives $R_{p_0}=0$, and Proposition~\ref{prop:second_order_cancellation}(ii) yields $|s_\delta(w)|\leq C\|w-p_0\|^3$ near $p_0$ (Fig.~\ref{fig:fl_phase}(d)). For any nonzero $v\in T_{p_0}\mathcal Z$, however, $v\in N_{p_0}\mathcal K\setminus\{0\}$, while $\mathrm d_{p_0}y_1[v]\in T_{p_0}\mathcal M$, $\mathrm d_{p_0}u_{2,i}[v]\in T_{p_0}\mathcal Z_i$, and $\mathrm d_{p_0}y_2[v]=\mathrm d_{p_0}u_1[v]=0$. Proposition~\ref{prop:quadratic-margin-obstruction} therefore excludes a quadratic margin $Q(w)\geq\alpha\|w-p_0\|^2$ on a full neighborhood of $p_0$, and the failure is genuine: since $\chi(p_0)=0$, initializations with $\chi_0\neq0$ converge to $p_{\chi_0}\neq p_0$, so $\dist(y_1(t),\mathcal M)\to\|p_{\chi_0}\|-\sqrt{2m}>0$ (Fig.~\ref{fig:fl_phase}(e)). On the balanced restriction $\mathcal H=\{\chi=0\}$, whose normal $\nabla\chi(p_0)$ spans $T_{p_0}\mathcal Z$, transversality gives $\|w-p_0\|\leq2\|e\|$ near $p_0$. Then~\eqref{eq:fl_gradient_bound} yields~\eqref{eq:quadratic_Q_margin} with $\alpha=\tfrac14\varepsilon_2 c_{\mathcal Z}$ on $\{\chi=0\}$ near $p_0$, where $T_{p_0}\{\chi=0\}=N_{p_0}\mathcal M$ ensures compact local storage sublevel components, and Corollary~\ref{cor:cubic_residual_absorption} gives $w(t)\to p_0$.

\subsection{Larger Circle: Nonzero Quadratic Residuals}\label{ssec:fl_incompatible}
For $\mathcal M=\mathcal C_\varrho$ with $\varrho>\sqrt{2m}$, the circle crosses $\mathcal Z$ transversally within $\mathcal P$, so $T_p\mathcal M\not\subseteq T_p\mathcal Z$ at each feasible point $p\in\mathcal Z_{\mathrm{cl}}=\mathcal M\cap\mathcal Z$. Proposition~\ref{prop:fl_heterogeneity}(ii) therefore gives $R_p\neq0$ (Fig.~\ref{fig:fl_phase}(d)), and the quadratic residual must be compared with the available dissipation. To apply Theorem~\ref{thm:net_dissipation_margin} via~\eqref{eq:available_dissipation} without the conservatism of $Q$, we retain the unrelaxed client supply $W_0(w):=\sum_i\langle e_{\mathcal Z_i}(w),\nabla\ell_i(w)\rangle\geq0$, for which~\eqref{eq:fl_defect} gives $W_0(w)-s_\delta(w)=\langle e_{\mathcal M}(w),\nabla L(w)\rangle=-\dot V_{\mathrm{cl}}(w)$. Since $e_{\mathcal M}(p)=\nabla L(p)=0$ and $\mathrm d_pe_{\mathcal M}=\mathrm P_{N_p\mathcal M}$, the net quadratic form~\eqref{eq:net_quadratic_form} at $p$ is
\begin{equation}\label{eq:fl_Hp}
H_p=\tfrac12\nabla^2W_0(p)-R_p=\operatorname{sym}\bigl(\mathrm P_{N_p\mathcal M}\nabla^2L(p)\bigr).
\end{equation}

For the data in Fig.~\ref{fig:fl_phase}, $X=\operatorname{diag}(9/4,5/4)$, $m=1$, and $\varrho=9/5$. At the feasible equilibrium $p=(a_p,\theta^\star/a_p)$ with
$a_p^2=\frac{81+\sqrt{4061}}{50}$, $a_p>0$, and $\chi(p)=\frac{\sqrt{4061}}{25}$,
an orthonormal basis of $T_p\mathcal H$ for the matched restriction $\mathcal H=\{\chi=\chi(p)\}$ is
$$
E=\begin{bmatrix}
1/(a_p\varrho)&0\\
4a_p/(5\varrho)&-3/5\\
3a_p/(5\varrho)&4/5
\end{bmatrix}.
$$
Direct calculation from~\eqref{eq:fl_Hp} yields
$$
\begin{aligned}
E^\top\mathrm P_{N_p\mathcal M}E
&=\operatorname{diag}(2500/6561,1)\succ0,\\
E^\top H_pE
&=\begin{bmatrix}
\tfrac73&-\tfrac{18122a_p}{30375}\\
-\tfrac{18122a_p}{30375}&\tfrac{161a_p^2}{100}
\end{bmatrix}\succ0,
\end{aligned}
$$
while $(E^\top R_pE)_{11}=28427/7500>0$, so the quadratic residual remains nonzero on $T_p\mathcal H$. Since $\nabla^2V_{\mathrm{cl}}(p)=\mathrm P_{N_p\mathcal M}$, the two positive-definite matrices verify~\eqref{eq:leaf_storage_comparison} and~\eqref{eq:restricted_net_positivity} on $\mathcal H$ with $\mathcal K=\{p\}$. Theorem~\ref{thm:net_dissipation_margin}(iii) therefore yields local exponential convergence to $p$ on $\mathcal H$ (Fig.~\ref{fig:fl_phase}(e), orange solid). If $\chi_0\notin\{\pm\sqrt{\varrho^4-4m^2}\}$, then $\{\chi=\chi_0\}\cap\mathcal Z_{\mathrm{cl}}=\emptyset$: under the local positive-branch hypotheses of Subsection~\ref{ssec:fl_exact}, $w(t)\to p_{\chi_0}\notin\mathcal C_\varrho$ and $\dist(y_1(t),\mathcal C_\varrho)\to\bigl|\|p_{\chi_0}\|-\varrho\bigr|>0$ (Fig.~\ref{fig:fl_phase}(e), orange dash-dotted).

\subsection{Discussion}\label{ssec:fl_discussion}
Since the clients are neither passive nor monotone, passivity- and monotonicity-based analyses of networked optimization~\cite{arcak2007passivity,burger2014duality} do not apply, and convergence analyses of overparameterized learning~\cite{liu2022loss,liu2023aiming} certify convergence to $\mathcal Z$ only. Transversal passivity, the hierarchy of Table~\ref{tab:residual_hierarchy}, and conserved-quantity confinement certify the same closed loop relative to other targets, identify through the imbalance which targets are reachable, and explain through the margin obstruction when certification fails. Heterogeneity enters only through the tangency condition of Proposition~\ref{prop:fl_heterogeneity}(ii), which is checked from the targets alone.

\section{Conclusion}\label{sec:conclusion}
We introduced transversal passivity, a storage inequality in the transverse port errors with respect to prescribed manifolds, and characterized it by a necessary condition in linear matrix inequality form, a constructive sufficient condition, and its relation to standard, shifted, and incremental passivity. For feedback through negative-adjoint linear channels, the non-cancelling supply is governed by the compatibility defects of the channels, which yields a hierarchy from exact compatibility through quadratic cancellation to nonzero quadratic residuals. Together with compact confinement and a dissipation margin, each level of this hierarchy gives port convergence. A federated learning example with heterogeneous data showed that convergence can be certified although no server target is compatible, and that a conserved quantity of the learning dynamics determines which targets are reachable.

Future work includes nonlinear interconnections, the synthesis of transversally passive systems, the role of dynamical and interconnection symmetries, and networked systems on directed graphs, whose channels need not form a negative-adjoint pair~\cite{yue2025Spassivity,yue2025EJC}, with decentralized learning as a particular application.

\section*{Acknowledgment}
This work was supported by the Israel Science Foundation grant no.\,453/24. The authors used Claude (Anthropic) and ChatGPT (OpenAI) to assist with copy-editing, with generating the code for the figures, and with the numerical studies in Section~\ref{sec:fl}. All AI-assisted content was reviewed and verified by the authors, who take full responsibility for this article.

\bibliographystyle{IEEEtran}
\bibliography{references}

\end{document}